\documentclass[reprint,amsmath,amssymb,aps,onecolumn]{revtex4-2}

\usepackage{graphicx}
\usepackage{dcolumn}
\usepackage{bm}
\usepackage{hyperref}
\usepackage[english]{babel}
\usepackage{xcolor}

\usepackage{color}

\newcommand{\e}{\varepsilon}
\newcommand{\w}{\omega}

\newcommand{\kp}{\kappa}

\newcommand{\ee}{\mathrm{e}}
\newcommand{\ii}{\mathrm{i}}
\newcommand{\dd}{\,\mathrm{d}}

\begin{document}

\preprint{APS/123-QED}

\title{
Cluster dynamics in a two-group Stuart-Landau model analyzed by the second-order phase reduction
}

\author{Yernur Baibolatov}
\email{yernurb@googlemail.com}
\affiliation{Department of Physics, Kazakh National Women's Teacher Training University, Aiteke Bi str.~99,
050040 Almaty, Kazakhstan}
\author{Oleh E. Omel’chenko}
\email[]{omelchenko@uni-potsdam.de}
\affiliation{Department of Physics and Astronomy, University of Potsdam, 
Karl-Liebknecht-Str. 24/25, D-14476 Potsdam-Golm, Germany}

\author{Michael Rosenblum}
\email[]{mros@uni-potsdam.de}
\affiliation{Department of Physics and Astronomy, University of Potsdam, 
Karl-Liebknecht-Str. 24/25, D-14476 Potsdam-Golm, Germany}

\date{\today}

\begin{abstract}
We analyze cluster states in an ensemble of Stuart-Landau oscillators
with two subpopulations of different frequencies. Our main goal is to
compare the descriptions of the system's dynamics obtained via the
standard first-order phase approximation and the second-order phase
reduction. We demonstrate that the second-order model not only
provides quantitative improvements in the description but also reveals
new dynamical states not present in the standard Kuramoto theory. In
particular, it describes bistability of synchronous states in the
minimal setup of two coupled oscillators and the existence of
three-cluster states, forbidden in the first-order phase description by the Watanabe-Strogatz theory.
The very good agreement between the second-order approximation results and
the results of numerical simulations of the original Stuart-Landau network
highlights the usefulness of high-order phase-reduction models.
\end{abstract}


\maketitle


{\bf
Phase approximation is a widely used standard tool for analyzing
weakly coupled oscillatory networks. Typically, phase oscillator
models are derived via first-order phase reduction; for a network of
interacting Stuart-Landau oscillators, this yields the celebrated
Kuramoto model. Many recent studies have sought to extend this
approach by deriving high-order phase oscillator models as power
series in the coupling strength. Here, we use the second-order model
to analyze cluster states in a system of coupled Stuart-Landau units;
the population consists of two groups of equal size, each with its own
frequency. We demonstrate that, under certain conditions, when the
interaction in the first-order approximation vanishes, 
accounting for quadratic terms becomes crucial and reveals new effects.
For example, in the simplest case of
two interacting oscillators, a bistability between in-
and anti-phase synchronous states occurs, and its analytical description fully agrees with the
numerical simulation for the original Stuart-Landau model. Moreover, only exploiting the second-order approximation, we can explain the three-cluster states in which the fastest
group splits into two clusters, while the slow one forms one. Such
states do not exist in the first-order Kuramoto model - they are
forbidden by the Watanabe-Strogatz theory - but are present in the
Stuart-Landau network. With these examples, our study demonstrates the
usefulness of high-order phase reduction}.

\section{Introduction}

Phase dynamics models are important mathematical tools for studying synchronization phenomena in oscillatory networks, with the celebrated Kuramoto model being the most popular setup~\cite{Kuramoto-75,kuramoto1984}. Derived half a century ago, this model and its various modifications remain the focus of intensive current research. In particular, many studies extend the model to account for multibody interaction \cite{battiston2020,battiston2021,battiston2022,NijO-EEKP2022,boccaletti2023,bick2023a,Gao2023HigherOrder,Millan2025}.
Among the different studies on this topic
\cite{skardal2020,LanR2022,AdhRS2023,SkaAR2023,bick2023c,Bick_2024,SumJ2024},
we are primarily interested in the models that arise from a phase-reduction process
\cite{wilson2019c,leon2019,wilson2020a,gengel2021,kumar2021,leon2022,mau2023c,nicks2024,bick2024a,Mau-Omelchenko-Rosenblum-24,leon2025}.

Y. Kuramoto developed his now-classical model by applying the phase-reduction technique to obtain a simplified description of a population of weakly interacting Stuart-Landau (SL) oscillators as the first-order approximation in the coupling strength. For many years, the higher-order approximations remained a theoretical challenge;
only recently have many groups tackled this problem,
going beyond the leading order. 
An important contribution in this direction was the explicit second-order phase-dynamics model of a network of SL oscillators, derived by Le\'on and Paz\'o~\citep{leon2019,leon2022}, which was later generalized to the case of arbitrary linear coupling~\citep{Mau-Omelchenko-Rosenblum-24}.
The most interesting features of these models are the appearance of multi-body interaction terms and pairwise interaction terms for non-connected units, as well as the frequency dependence of the terms' coefficients. The significance of the second-order phase approximation was demonstrated through examples of a more accurate description of synchronization transitions~\cite{leon2019}, and an adequate representation of a transition
to collective chaos in the SL oscillators model \citep{leon2022}.
It proved to be useful for calculating dynamics corrections due to the non-circular shape of the limit cycle~\cite{bick2024a} 
and for a deeper understanding of
pattern formation on networks, including the explanation of remote synchrony~\cite{kumar2021}
and the description of the chimera's 
shape~\cite{Mau-Omelchenko-Rosenblum-24}. 
Of particular interest are qualitative results when the second-order phase equations describe effects not captured by the traditional Kuramoto model.  

In this paper, we present another example of a second-order analysis of an oscillatory network. We study the dynamics of a two-group population of SL units and demonstrate that accounting for second-order terms not only quantitatively improves the system's description but also reveals new effects not captured in the first-order phase approximation. Remarkably, the second-order model allows for an analytical treatment. As a result, we obtain a detailed description of different cluster states, their stability, and synchronization transitions involving them.

The paper is organized as follows. In Section II, we present the two-group Stuart-Landau model and its second-order phase approximation.
In Section III, we analyze the simplest setup of two coupled oscillators. Next, Section IV presents the analysis of the two- and the three-cluster states. In Section V, we discuss our results. Some technical details are given in Appendices.

\section{The model}
In this paper, we consider a popular model of Stuart-Landau oscillators with global diffusive coupling~\citep{HakimRappel1992,NakagawaKuramoto1993,NakagawaKuramoto1994,DaidoNakanishi2006,Aranson2002,GarciaMorales2012}. As is well known, the SL equation is a normal form for the Andronov-Hopf bifurcation; thus, the SL network is a universal description of coupled self-sustained systems close to the oscillation onset point.
SL networks have been extensively studied
in their original form~\citep{Matthews1991,HakimRappel1992,DaidoNakanishi2006} as well as in phase approximation,
exploiting the celebrated Kuramoto and Kuramoto-Sakaguchi models~\citep{sakaguchi1986}
and their more elaborated versions~\citep{leon2019,leon2022}.
Our model reads:
\begin{align}
    \dot z_i = (\eta +  \ii \nu_i)z_i - \eta |z_i|^2 z_i+\frac{\e \ee^{\ii \alpha}}{N}  \sum_{j=1}^N ( z_j-z_i)\,,
    \label{SL:general}
\end{align}
where $z_i\in\mathbb{C}$ is the state variable of the $i$th oscillator and $\nu_i$ are real-valued parameters
that determine individual frequencies.
To minimize the number of parameters, we consider isochronous oscillators; however, we take a complex-valued coupling strength $\e\exp{(\ii\alpha)}$. (Notice that in the first-order phase approximation the oscillator's non-isochronicity and coupling parameter $\alpha$ play a similar role, introducing a constant phase shift in the sine coupling function.)

In the following, we focus on the special case
where the parameters $\nu_i$ are selected
from a bimodal delta-distribution.
More precisely, we assume $\nu_i=\w_1$, for $i=1,\ldots,N_g$,
and $\nu_i=\w_2$, for $i=N_g+1,\ldots,2N_g$,
with the total number of oscillators $N=2N_g$.
Then, system~(\ref{SL:general}) can be rewritten in the form
\begin{equation}
\dot{z}_i^{(n)} = ( \eta + \ii \omega^{(n)} ) z_i^{(n)} - \eta \left|z_i^{(n)}\right|^2 z_i^{(n)}
+ \frac{\e \ee^{\ii \alpha}}{2 N_\mathrm{g}} \sum\limits_{l=1}^2 \sum\limits_{j=1}^{N_\mathrm{g}}  \left( z_j^{(l)} - z_i^{(n)} \right),
\label{SL_2groups}
\end{equation}
where $n=1,2$ and $i=1,\dots,N_\mathrm{g}$.
Without loss of generality, we assume that the coupling parameters are $\e>0$ and $0<\alpha<\pi$ 
(with an exceptional case $\alpha=0$ when $\e$ can be both positive or negative).

Using the general result of~\cite{Mau-Omelchenko-Rosenblum-24}, we write the second-order phase-reduced model
corresponding to Eq.~(\ref{SL_2groups}): 
\begin{eqnarray}
\dot{\phi}_i^{(n)}&=& \omega^{(n)} - \e \sin\alpha + \frac{\e}{2 N_\mathrm{g}}
\sum\limits_{l=1}^2 \sum\limits_{j=1}^{N_\mathrm{g}}
\sin\left( \phi_j^{(l)} - \phi_i^{(n)} + \alpha \right) \nonumber\\[2mm]
&+& \frac{\e^2}{8 \kappa N_\mathrm{g}^2}
\sum\limits_{l=1}^2 \sum\limits_{j=1}^{N_\mathrm{g}}
\sum\limits_{m=1}^2 \sum\limits_{k=1}^{N_\mathrm{g}} \left\{
\cos(\Delta_{nm}) \sin\left( \phi_j^{(l)} - \phi_k^{(m)} + \Delta_{nm} \right) \right. \nonumber\\[2mm]
&+& \cos(\Delta_{nm}) \sin\left( \phi_k^{(m)} + \phi_j^{(l)} - 2 \phi_i^{(n)} - \Delta_{nm} + 2\alpha \right) \nonumber\\[2mm]
&-& \cos(\Delta_{lm}) \sin\left( 2 \phi_j^{(l)} - \phi_k^{(m)} - \phi_i^{(n)} + \Delta_{lm} \right)\nonumber \\[2mm]
&-& \left. \cos(\Delta_{lm}) \sin\left( \phi_k^{(m)} - \phi_i^{(n)} - \Delta_{lm} + 2\alpha \right) \right\},
\label{phase_eq_with_delta}
\end{eqnarray}
where $\Delta_{nm} = \arctan[ ( \omega^{(n)} - \omega^{(m)} ) / \kappa ]$
and $\kappa = - 2 \eta$ is the Floquet exponent quantifying the stability of the autonomous oscillator's limit cycle. Alternatively, 
Eq.~(\ref{phase_eq_with_delta}) can be written in a concise form
\begin{eqnarray}
\dot{\phi}_i^{(n)} &=& \omega^{(n)} - \e \sin\alpha + \frac{\e}{2}
\sum\limits_{l=1}^2 \mathrm{Im}\left[ Z^{(l)} \ee^{\ii \alpha} \ee^{-\ii \phi_i^{(n)}}  \right]
+ \frac{\e^2}{8 \kappa} \sum\limits_{l=1}^2 \sum\limits_{m=1}^2 \left\{
\mathrm{Im}\left[ Z^{(l)} \overline{Z}^{(m)} \zeta_{nm} \right] \right. \nonumber\\[2mm]
&+& \left. \mathrm{Im}\left[ Z^{(m)} Z^{(l)} \overline{\zeta}_{nm} \ee^{2 \ii \alpha} \ee^{- 2 \ii \phi_i^{(n)} } \right]
- \mathrm{Im}\left[ Z_2^{(l)} \overline{Z}^{(m)} \zeta_{lm} \ee^{- \ii \phi_i^{(n)}} \right]
- \mathrm{Im}\left[ Z^{(m)} \overline{\zeta}_{lm} \ee^{2 \ii \alpha} \ee^{- \ii \phi_i^{(n)} } \right] \right\}\;,
\label{phase_eq_general}
\end{eqnarray}
where we use for brevity 
$\zeta_{nm} = \cos(\Delta_{nm}) \ee^{\ii \Delta_{nm}}$ and introduce the mean fields
\begin{equation}
Z^{(m)} = \frac{1}{N_\mathrm{g}} \sum\limits_{j=1}^{N_\mathrm{g}} \ee^{\ii \phi_j^{(m)}},
\quad
Z_2^{(m)} = \frac{1}{N_\mathrm{g}} \sum\limits_{j=1}^{N_\mathrm{g}} \ee^{2 \ii \phi_j^{(m)}}\,.
\label{complex_mean_fields}
\end{equation}

Note that typically
$|\omega^{(n)} - \omega^{(m)}| \ll |\kappa|$ so that
$\Delta_{nm}\approx 0$ and $\zeta_{nm}\approx 1$. 
For such a case, we further simplify the phase model in accordance with the results of Le\'on and Paz\'o~\cite{leon2019}:
\begin{eqnarray}
\dot{\phi}_i^{(n)} &=& \omega^{(n)} - \e \sin\alpha + \frac{\e}{2}
\sum\limits_{l=1}^2 \mathrm{Im}\left[ Z^{(l)} \ee^{\ii \alpha} \ee^{-\ii \phi_i^{(n)}} \right] \nonumber\\[2mm]
&+& \frac{\e^2}{8 \kappa} \sum\limits_{l=1}^2 \sum\limits_{m=1}^2\left\{ \mathrm{Im}\left[ Z^{(m)} Z^{(l)} \ee^{2 \ii \alpha} \ee^{- 2 \ii \phi_i^{(n)} } \right]
- \mathrm{Im}\left[ Z_2^{(l)} \overline{Z}^{(m)} \ee^{- \ii \phi_i^{(n)}} \right]
- \mathrm{Im}\left[ Z^{(m)} \ee^{2 \ii \alpha} \ee^{- \ii \phi_i^{(n)} } \right] \right\}.
\label{phase_eq_Leon}
\end{eqnarray}
In the following, we will consider the system~(\ref{phase_eq_Leon}).

\section{The simplest case: two oscillators}

In this section, we consider the simplest case of model~(\ref{SL_2groups}), 
where each population consists of a single oscillator, i.e,  $N_\mathrm{g} = 1$. 
We simplify the notations: 
using $\phi_{1,2}$ and $\omega_{1,2}$ instead of $\phi_1^{(n)}$ and $\omega^{(n)}$ and expanding the sums, we explicitly write:

\begin{equation}
    \dot{\phi}_1 = \omega_1 - \frac{\e}{2}\sin\alpha - \frac{\e}{2}\sin\left(\phi_{1,2}-\phi_{2,1}-\alpha\right) 
    + \frac{\e^2}{8\kappa}\biggl(-\sin(2\alpha)
    +\sin(2\phi_{1,2}-2\phi_{2,1})-\sin(2\phi_{1,2}-2\phi_{2,1}-2\alpha)\biggr)\,.
    \label{eq:two_phases}
\end{equation}
Introducing $\psi=\phi_1-\phi_2$ and 
$\delta=\w_1-\w_2$, and subtracting equations for $\dot\phi_{1,2}$, we obtain
\begin{equation}
\dot\psi=\delta-\e\cos\alpha\sin\psi+\frac{\e^2}{4\kp} \biggl(1-\cos(2\alpha) \biggr)\sin(2\psi)=\delta+f(\psi)\;,
\label{Adler}
\end{equation}
where the quadratic in $\e$ term represents the second-order correction to the well-known 
Adler equation. 

To introduce some notations, we recall the solution of the Adler equation 
$\dot\psi=\delta-\e\cos\alpha\sin\psi$. The synchronous solution $\psi=\text{const}$ exists
within the triangular Arnold tongue; its semi-width for some coupling strength $\e$ is 
$D= |\e\cos\alpha|$. The value of the constant phase shift $\psi$ depends on the 
detuning $\delta$. For attractive coupling, $\e\cos\alpha>0$, one has 
$-\Psi\le\psi\le\Psi$, with the maximal phase shift $\Psi=\pi/2$ (the in-phase solution). 
For repulsive coupling, $\e\cos\alpha<0$, one has 
$\pi-\Psi\le\psi\le\pi+\Psi$ (the anti-phase solution). 
As shown below, the tongue's semi-width and maximal phase shift generally differ for the in-phase and anti-phase states; therefore, we use the notations $D_{\mathrm{i,a}}$ and $\Psi_{\mathrm{i,a}}$ in the following. We will also see that in the second-order approximation,  $\Psi_{\mathrm{i,a}}$ is generally not constant but depends on $\e,\alpha$.

Let us start the analysis of the second-order Adler equation by considering 
two special cases.

\begin{itemize}
\item Case 1, $\alpha=0$. The second-order correction vanishes, Eq.~(\ref{Adler}) reduces to 
\[
\dot\psi=\delta-\e\sin\psi\;.
\]
Depending on the sign of $\e$, we have either an in-phase or an anti-phase solution.
The tongue's semi-width is $D_{\mathrm{i}}=D_\mathrm{a}=|\e|$ and 
$\Psi_\mathrm{i}=\Psi_\mathrm{a}=\pi/2$.
\item Case 2, $\alpha=\pi/2$. Now, the first-order term vanishes, meaning that in the first approximation, the coupling is neutral and there is no synchrony for any $\e$.  
The analysis of the second-order equation 
\[
\dot\psi=\delta +\frac{\e^2}{2\kp} \sin(2\psi)\;
\]
is simple and yields the synchrony condition $|\delta|\le \frac{\e^2}{2|\kp|}$ (we recall 
the Floquet exponent $\kappa<0$). An 
interesting feature is the appearance of bistability: in the synchronous state, 
$\sin(2\psi)=2|\kappa|\delta/\e^2$; if some $\psi_0$ satisfies this equation, 
$\psi_0+\pi$ is the root, too.
Thus, the in-phase states $-\pi/4\le\psi\le \pi/4$ and anti-phase states 
$\pi-\pi/4\le\psi\le \pi+\pi/4$ coexist.
Hence, $D_\mathrm{i}=D_\mathrm{a}=\e^2/2|\kappa|$ and the phase shift variation parameter 
$\Psi_\mathrm{i}=\Psi_\mathrm{a}=\pi/4$.
\end{itemize}

\begin{figure}[!htb]
\centering
\includegraphics[width=0.6\columnwidth]{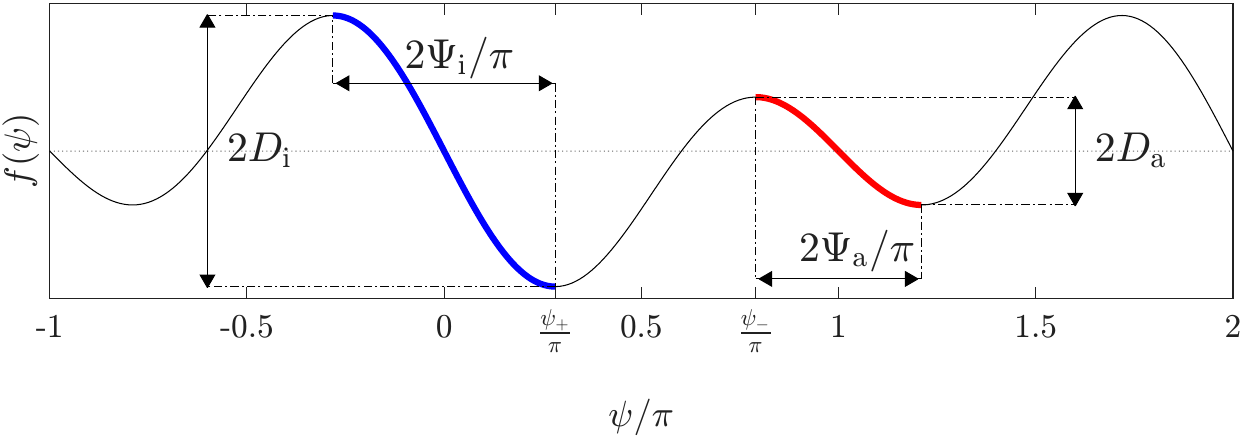}
\caption{An example of $f(\psi)$ (1.5 periods are shown) for the case of bistability.
Blue and red colored intervals determine the domains with in-phase and anti-phase locking, respectively. The difference of endpoints' ordinates of these domains determines the width of the corresponding tongues, while the difference of endpoints' abscissae determines the span of $\psi$ variation across the tongue.
} 
\label{Ffunc}
\end{figure}

To determine the synchronization domains for general $0<\alpha<\pi$, we have to find 
extrema of the function $f(\psi)$, defined in Eq.~(\ref{Adler}) and exemplified in 
Fig.~\ref{Ffunc}. (Notice that due to symmetry, $f_{\mathrm max}=-f_\mathrm {min}$.)  Thus, we have to solve
\begin{equation}
f'=-\e\cos\alpha\cos\psi-\frac{\e^2}{2\kp} \cos(2\psi)   \biggl(\cos(2\alpha)-1   \biggr)
=0\;.
\label{eq:der}
\end{equation}
With account of $\cos(2\psi)=2\cos^2\psi-1$, this yields a quadratic equation for $\cos\psi$; existence of two roots means bistability.
We rewrite Eq.~\ref{eq:der} as
$2\cos^2\psi+c\cos\psi-1=0$, 
with 
\begin{equation}
c=\frac{|\kp|\cos\alpha}{\e\sin^2\alpha}=\frac{\cos\alpha}{e\sin^2\alpha}\;,
\label{eq:c}
\end{equation}
where, for brevity, we introduced $e=\e/|\kp|$; the relevant parameter, as expected, 
is the ratio of the coupling strength and the measure of the cycle's stability.
The solution of the quadratic equation is:
\begin{equation}
\cos\psi_\pm=\frac{-c\pm\sqrt{c^2+8}}{4}\;,
\label{eq:cospm}
\end{equation}
where $\cos\psi_+>0$ (this root corresponds to the synchronous in-phase state) and 
$\cos\psi_-<0$ (anti-phase solution), cf. Fig.~\ref{Ffunc}.
The condition $|\cos\psi_+|\le 1$ yields $c\ge -1$, while the condition $|\cos\psi_-|\le 1$
yields $c\le 1$.  
Now, we re-formulate these existence conditions in terms of $\alpha$, first, for the root $\cos\psi_+$.  
Since $c\ge -1$, $\cos\alpha$ can be negative; thus, contrary to the first-order 
approximation, the synchronous in-phase solution exists for 
$0\le\alpha\le \alpha_\mathrm{i}$, 
where $\alpha_i$ we find from the condition $\cos\alpha_\mathrm{i}=e(\cos^2\alpha_\mathrm{i}-1)$ (see Eq.~(\ref{eq:c})).
Similarly,  the anti-phase solution exists not only for negative $\cos\alpha$, but, contrary to the 
first-order approximation, also for $\alpha_\mathrm{a}\le \alpha<\pi/2$. We find $\alpha_\mathrm{a}$ from the 
equation $\cos\alpha_\mathrm{a}=e(1-\cos^2\alpha_\mathrm{a})$.
The solution of these quadratic equations is
 (the second root is larger than one): 
\[
\cos\alpha_{\mathrm{i,a}}=\mp\frac{\sqrt{4\,e^2+1}-1}{2\,e}\;.
\]
Notice that 
$\alpha_\mathrm{i}=\pi-\alpha_\mathrm{a}>\alpha_\mathrm{a}$ which means coexistence of the in-phase and anti-phase solutions
in the interval $[\alpha_\mathrm{a},\alpha_\mathrm{i}]$.

\begin{figure}[!htb]
\centering
\includegraphics[width=\columnwidth]{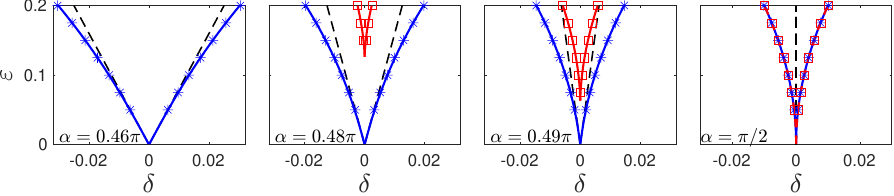}
\caption{Arnold tongues for two coupled oscillators. Dashed lines show the first-order solution, while blue and red curves present the tongues 
for the in-phase and anti-phase solutions, respectively. 
Blue stars and red squares correspond to the numerical simulation of Eq.~(\ref{SL_2groups}) for $N_g=1$.
The picture is symmetric with respect to $\alpha$: e.g., the plot for $\alpha=0.51\pi$ looks
like the plot for $\alpha=0.49\pi$, but blue and red colors interchange.
} 
\label{Tongues}
\end{figure}

\begin{figure}[!htb]
\centering
\includegraphics[width=0.6\columnwidth]{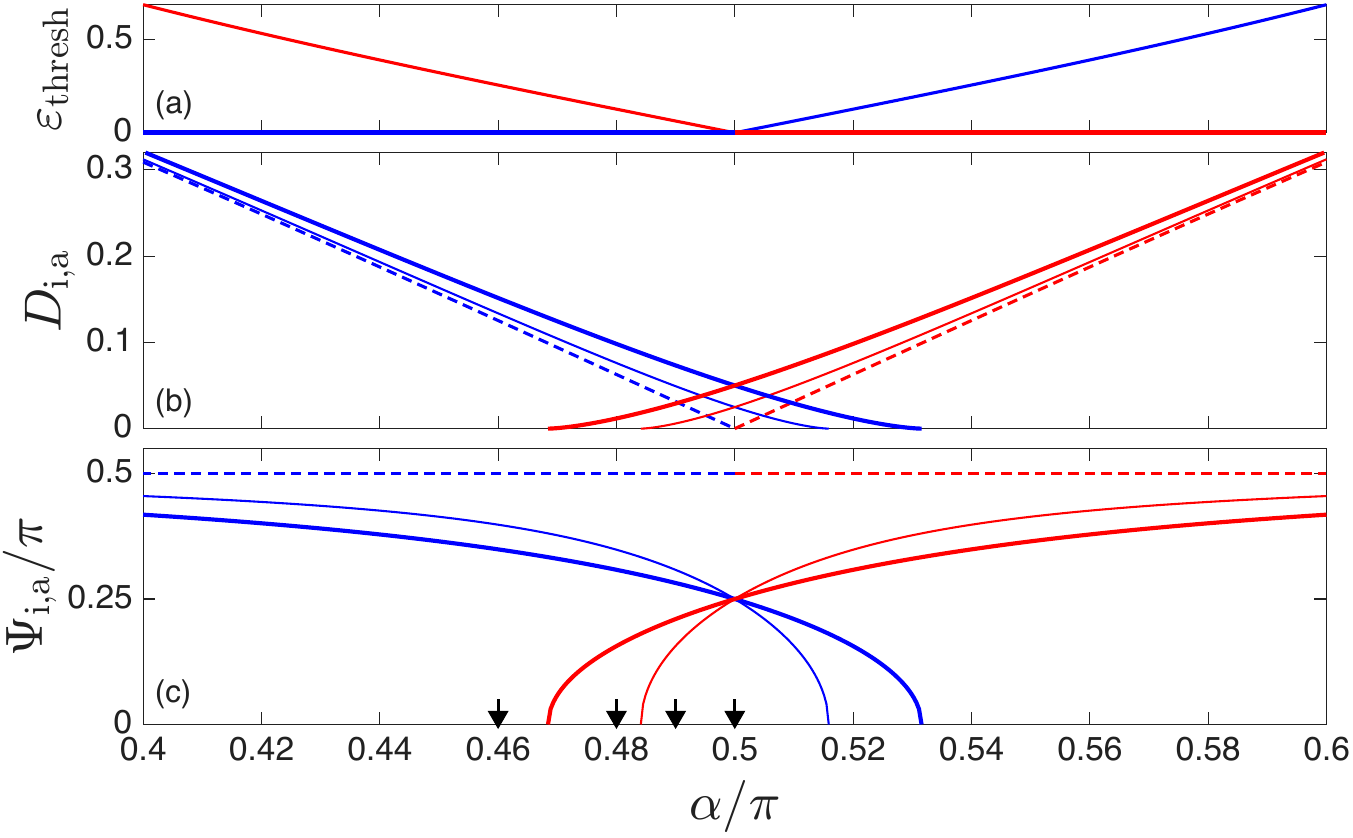}
\caption{Panel (a) shows the threshold coupling value of the Arnold tongue (the height of the tongue's tip) as a function of $\alpha$. 
In all panels, blue and red curves illustrate
the synchronous in-phase and anti-phase solutions, respectively. 
In panels (b,c), the solid and bold curves correspond 
to $\e=0.1$ and $\e=0.2$.
The dashed line shows the first-order solution, 
that is $|\cos\alpha|$ for 
$D_{i,a}/\e$ (b) and $\Psi_{\mathrm{i,a}}=\pi/2$ (c).
Notice that for the second-order model $\Psi_{\mathrm{i,a}}$ depends on $\e$ and $\alpha$ 
and $\Psi_\mathrm{i}(\alpha=\pi/2)=\Psi_\mathrm{a}(\alpha=\pi/2)=\pi/4$ for any $\e$.
For the values of $\alpha$ where the semi-width of the in-phase (blue) and anti-phase (red)
synchronization domains $D_{\mathrm{i,a}}>0$, the system is bistable; the width of the bistability region 
grows approximately proportional to $\e$.
Four black arrows indicate the $\alpha$-values corresponding to four panels in Fig.~\ref{Tongues}.
} 
\label{D_Psi}
\end{figure}

Now, we find the borders of each tongue: the tongue's semi-width for the 
in-phase and anti-phase locking is $D_\mathrm{i}=|f(\psi_+)|$ and $D_\mathrm{a}=|f(\psi_-)|$, respectively.
Fig.~\ref{Tongues} presents tongues for four different values of $\alpha$: 
$0.46\pi$, $0.48\pi$, $0.49\pi$, and $0.5\pi$. We see that the synchronous anti-phase solution exists 
for coupling strength $\e>\e_\mathrm{thresh}$.
We find this critical value from the condition $c=1$, which yields 
$\e_\mathrm{thresh}=|\kp| \cos\alpha/\sin^2\alpha$.
We see that this threshold is zero for $\alpha=\pi/2$ and tends to infinity with 
$\alpha\to 0$ and $\alpha \to \pi$. 
$\e_\mathrm{thresh}$ grows rapidly with the deviation of $\alpha$ from $\pi/2$, see Fig.~\ref{D_Psi}a. So, for $\alpha=0.4\pi$ it is about $0.7$ --- this value is definitely above the range of validity of the phase approximation. Thus, in practice, bistability exists over a relatively narrow interval around $\alpha=\pi/2$. Fig.~\ref{D_Psi}b provides another illustration of the bistability. 
Note that in this work, for all examples with numerical simulations of Eq.~(\ref{SL_2groups}), we use the parameter values $\omega_1=1$ and $\eta=1$.

The next important characteristic is $\Psi$, i.e., the range of $\psi$ variation across the tongue. 
In the first approximation, $\Psi_\mathrm{i}=\Psi_\mathrm{a}=\pi/2$ for any 
$\alpha$ (for $\alpha=\pi/2$ the tongue vanishes). 
In the second approximation, for  $\alpha=\pi/2$ we have $\Psi_\mathrm{i}=\Psi_\mathrm{a}=\pi/4$, 
for any $\e$. 
For arbitrary $\alpha$ and $\e$, we obtain 
\begin{equation}
\Psi_\mathrm{i}=\psi_+\quad \text{and} \quad    \Psi_\mathrm{a}=\pi-\psi_-
\label{eq:Psi_i:Psi_a}
\end{equation}
 for the in-phase and the anti-phase solution, respectively.
With the help of Eqs.~(\ref{eq:c},\ref{eq:cospm})
we compute $\Psi_{\mathrm{i,a}}$ explicitly and present the 
results in Fig.~\ref{D_Psi}c for $\kp=-2$ and $\e=0.1$, $\e=0.2$.

\section{Cluster states}
So far, we have analyzed the simplest case of two coupled oscillators. We now turn to the general case $N_g \gg 1$, 
where each population contains many oscillators and, hence, we can expect the formation of clusters. Clustering is a well-known phenomenon observed both in phase approximations~\citep{Okuda-93,Ashwin2007} and full models~\citep{GolombRinzel1993,GolombRinzel1994,Pikovsky-Popovych-Maistrenko-01,Kori2014}, in particular, in SL networks~\citep{HakimRappel1992,NakagawaKuramoto1993,NakagawaKuramoto1994,DaidoNakanishi2007,Kemeth2019,Kemeth2021,Thome2025}. 
Our primary goal is to compare cluster configurations in our two-group SL network and in the corresponding first- and second-order approximations. 

Our system exhibits various cluster states, which we classify using the $(p+q)$ notation: the first group splits into $p$ clusters and the second into $q$, with all oscillators within each cluster sharing the same phase.
The most common regime is the $(1+1)$ state, where each group forms a single cluster. The entire ensemble of $2N_g$ 
oscillators then behaves as a system of two macroscopic oscillators, and the phase difference between the clusters 
obeys the same second-order Adler equation (\ref{Adler}) as in the two-oscillator case. Depending on the parameters,
the two clusters can be mutually synchronized or asynchronous. The crucial difference from the genuine 
two-oscillator system is that clusters can lose their stability and evaporate, i.e., individual oscillators can 
detach from the cluster, destroying the cluster state.

Beyond the $(1+1)$ state, we also observe the $(2+1)$ regime, where one (faster) group splits into two clusters 
while the other group remains synchronized. We suspect that other $(p+q)$ combinations are possible in 
principle, but have not observed them in our simulations. In all cases, we observed that the clusters within each 
group are of equal size, although our analysis does not rely on this property.

Before analyzing cluster stability, we emphasize that 
the $(2+1)$ cluster states in the first-order model 
are forbidden by the Watanabe-Strogatz (WS) theory~\citep{Watanabe-Strogatz-93,Watanabe-Strogatz-94,Pikovsky-Rosenblum-15}: for generic initial conditions, the first group cannot split into two clusters of equal size. 
According to the WS theory, the dynamics of this group of $N_g$ identical oscillators evolving under a common forcing obey three differential equations for collective variables $\rho$, $\Xi$, $\Gamma$, where $0\le\rho\le 1$ and $\Xi$, $\Gamma$ are angles. In addition, the WS equations contain $N_g$ angular constants of motion $\xi_i$, determined from initial conditions; these constants obey three additional constraints. 
For different initial conditions, all constants differ.
The oscillators' phases are retrieved from the collective variables by the M\"obius transformation
 \[
 \ee^{\ii\phi_i}=\ee^{\ii\Gamma}\frac{\rho+\ee^{\ii(\xi_i-\Xi)}}{\rho\ee^{\ii(\xi_i-\Xi)}+1}\;.
 \]
 For $\rho<1$ this transformation is one-to-one on the unit circle; hence, for different $\xi_i$ phases remain different
 (no clusters). For $\rho\to 1$, the transformation becomes singular and maps every value of $\xi_i$ to the same phase.
 Thus, only one cluster is possible for generic initial conditions. (An exceptional situation occurs if 
 $\xi_i-\Xi=\pi$ for some $i=j$; then oscillator $j$ may split from others. For a discussion of such solitary states, 
 see \citep{Maistrenko-Penkovsky-Rosenblum-14}.) However, below we demonstrate that the second-order approximation admits splitting of the group into two halves.

To analyze the cluster states, it is convenient to collect the terms with $\ee^{-\ii\phi_i^{(n)}}$ 
and rewrite Eq.~(\ref{phase_eq_Leon}) in the following form:
\begin{equation}
\dot{\phi}_i^{(n)} = \Omega^{(n)} + \mathrm{Im}\left[f_1\left(Z^{(1,2)},Z_2^{(1,2)}; \e, \alpha, \kappa\right)\cdot\ee^{-\ii\phi_i^{(n)}} 
    + f_2\left(Z^{(1,2)},Z_2^{(1,2)}; \e, \alpha, \kappa\right)\cdot\ee^{-2\ii\phi_i^{(n)}} \right],
\label{phase_eq_generic}
\end{equation}
where $f_1\left(Z^{(1,2)},Z_2^{(1,2)}; \e, \alpha, \kappa\right)$ and $f_2\left(Z^{(1,2)},Z_2^{(1,2)}; \e, \alpha, \kappa\right)$ 
are the functions of the mean fields only, and $\Omega^{(n)}$ are constants. In the $(p+q)$ state the phases of 
the oscillators of the first group are $\phi_i^{(1)}\in \{\Theta_1,\dots,\Theta_p\}$
and the phases of the oscillators of the second group are $\phi_i^{(2)}\in\{\Phi_1,\dots,\Phi_q\}$, where $\Theta_s$
and $\Phi_r$ are phases of the clusters of the first and second groups, respectively. Next, we can write 
equations of the form~(\ref{phase_eq_generic}) for each $\Theta_s$ and $\Phi_r$:
\begin{eqnarray}   
\dot{\Theta}_s &=& \Omega_1 + \mathrm{Im}\left[f_1\left(Z^{(1,2)},Z_2^{(1,2)}; \e, \alpha, \kappa\right)\cdot\ee^{-\ii\Theta_s} 
    + f_2\left(Z^{(1,2)},Z_2^{(1,2)}; \e, \alpha, \kappa\right)\cdot\ee^{-2\ii\Theta_s} \right],\qquad s=1,\dots,p,
\label{cluster_eq_generic_a} \\
\dot{\Phi}_r &=& \Omega_2 + \mathrm{Im}\left[f_1\left(Z^{(1,2)},Z_2^{(1,2)}; \e, \alpha, \kappa \right)\cdot\ee^{-\ii\Phi_r} 
    + f_2\left(Z^{(1,2)},Z_2^{(1,2)}; \e, \alpha, \kappa\right)\cdot\ee^{-2\ii\Phi_r} \right],\qquad r=1,\dots,q.
\label{cluster_eq_generic_b}
\end{eqnarray}

To assess the transversal stability of each cluster, i.e., whether individual oscillators remain locked to their
cluster or evaporate from it, we follow~\cite{Kaneko-94,Pikovsky-Popovych-Maistrenko-01}, and add a virtual 
test oscillator to each cluster. This oscillator is subject to the same forcing as all cluster elements but does not contribute to the mean field.
By linearizing Eqs.~\eqref{cluster_eq_generic_a}
and~\eqref{cluster_eq_generic_b}
for a small perturbation of the test oscillator,
we obtain the linear equations that determine the cluster's stability.
The only assumption underlying the stability analysis is that each cluster is sufficiently large so that the effect of a single oscillator's evaporation, which is of order $\sim N^{-1}$, has a negligible effect on the mean field.

More specifically, the phases of the test oscillators
can be written as $\theta_s = \Theta_s+\delta\theta_s$ and $\phi_r = \Phi_r+\delta\phi_r$,
where $\delta\theta_s$ and $\delta\phi_r$ denote the deviations from the dynamics of the corresponding clusters. For small 
$\delta\theta_s$ and $\delta\phi_r$, the exponent terms linearize as 
$\ee^{-\ii\delta\theta_s}\approx 1-\ii\delta\theta_s$ and $\ee^{-2\ii\delta\theta_s}\approx 1-2\ii\delta\theta_s$, and we obtain the instantaneous growth-rate factors
$$
    \frac{\dd\delta\theta_s}{\dd t} = \Lambda_s\left(Z^{(1,2)},Z_2^{(1,2)};\e,\alpha,\kappa\right)\cdot\delta\theta_s,\qquad
    \frac{\dd\delta\phi_r}{\dd t} = \Lambda_r\left(Z^{(1,2)},Z_2^{(1,2)};\e,\alpha,\kappa\right)\cdot\delta\phi_r,
$$
where 
\begin{eqnarray}
    \Lambda_s\left(Z^{(1,2)},Z_2^{(1,2)};\e,\alpha,\kappa\right) &=& 
    -\mathrm{Re}\left[f_1\left(Z^{(1,2)},Z_2^{(1,2)};\e,\alpha,\kappa\right)\cdot\ee^{-\ii \Theta_s}
    +2f_2\left(Z^{(1,2)},Z_2^{(1,2)};\e,\alpha,\kappa\right)\cdot\ee^{-2\ii \Theta_s}\right], 
    \label{TLE_general_form_a} \\
    \Lambda_r\left(Z^{(1,2)},Z_2^{(1,2)};\e,\alpha,\kappa\right) &=& 
    -\mathrm{Re}\left[f_1\left(Z^{(1,2)},Z_2^{(1,2)};\e,\alpha,\kappa\right)\cdot\ee^{-\ii \Phi_r}
    +2f_2\left(Z^{(1,2)},Z_2^{(1,2)};\e,\alpha,\kappa\right)\cdot\ee^{-2\ii \Phi_r}\right].
    \label{TLE_general_form_b}
\end{eqnarray}
The transversal Lyapunov exponents (TLEs)
of the $(p+q)$-cluster state are defined
as the average decay/growth rates
of the test oscillator perturbations; they are given by
\begin{equation}
\lambda_{s,r} = \lim\limits_{T\to\infty} \frac{1}{T}
\int_0^T \Lambda_{s,r}\left(Z^{(1,2)}(t),Z_2^{(1,2)}(t);\e,\alpha,\kappa\right) \dd t\;.
\label{averageTLE}
\end{equation}

Note that from the original phase-reduced equations~(\ref{phase_eq_Leon}), we have
\begin{eqnarray}
    \Omega_{1,2} &=& \omega_{1,2}-\e\sin\alpha, \\
    f_1\left(Z^{(1,2)},Z_2^{(1,2)}; \e,\alpha,\kappa\right) &=& \frac{\e}{2}\ee^{\ii\alpha}\left(1-\frac{\e}{2\kappa}\ee^{\ii\alpha}\right)\xi_1
    - \frac{\e^2}{8\kappa}\xi_2, \label{Def:f_1}\\
    f_2\left(Z^{(1,2)},Z_2^{(1,2)};\e,\alpha,\kappa\right) &=& \frac{\e^2}{8\kappa} \ee^{2\ii\alpha}\xi_1^2, \label{Def:f_2}
\end{eqnarray}
where
\begin{equation}
    \xi_1 = Z^{(1)}+Z^{(2)}, \qquad 
    \xi_2 = \left(\overline{Z}^{(1)}+\overline{Z}^{(2)}\right)\left(Z_2^{(1)}+Z_2^{(2)}\right).
\label{Def:xi_1:xi_2}
\end{equation}
Therefore, to analyze the state with a specific cluster configuration, we only need to express the mean 
fields $Z$, $Z_2$. In particular, for clusters of equal sizes within each group, the mean fields of a general $(p+q)$ state are given by
\begin{eqnarray}
    Z^{(1)} &=& \frac{1}{p}\sum_{s=1}^p\ee^{\ii\Theta_s},\qquad Z^{(2)} = \frac{1}{q}\sum_{r=1}^q\ee^{\ii\Phi_r},\\ 
    Z_2^{(1)} &=& \frac{1}{p}\sum_{s=1}^p\ee^{2\ii\Theta_s},\qquad 
    Z_2^{(2)} = \frac{1}{q}\sum_{r=1}^q\ee^{2\ii\Phi_r}. 
\label{mean_fields_cluster_state}
\end{eqnarray}


\subsection{Two clusters}
The simplest cluster state configuration is the two-cluster state, or $(1+1)$ state with $p=1$ and $q=1$. In this state,
the oscillators in each group synchronize; the phases fulfill $\phi_i^{(1)}=\Theta$ and $\phi_i^{(2)}=\Phi$, 
for all $i=1,\ldots,N_g$.
The cluster phases $\Theta$ and $\Phi$ obey Eqs.~(\ref{cluster_eq_generic_a}),(\ref{cluster_eq_generic_b}).
Treating each cluster like a macroscopic oscillator, 
we write, in analogy to Eq.~(\ref{Adler}), the second-order Adler equation for the phase shift 
$\Psi=\Theta-\Phi$:

\begin{equation}
    \dot{\Psi} = \delta - \e\sin\Psi \cos\alpha + \frac{\e^2}{4\kappa}\biggl( 1-\cos(2\alpha) \biggr)\sin(2\Psi)\,.     \label{Adler_equation_1+1}
\end{equation}
We emphasize the crucial difference from the two-oscillator case: Eq.~(\ref{Adler_equation_1+1}) is valid only as long as the clusters remain stable. 

Before proceeding with the analysis of the clusters' stability, we mention that since two clusters behave like two macroscopic oscillators, 
the same locking conditions apply.
Thus, for given $\e$, we find the range of frequency detuning $\delta$ such that the phase difference $\Psi$
remains constant. 
(Similarly to the two-oscillator case, we write for this constant 
$-\Psi_\mathrm{i}^{(c)}\le \Psi \le \Psi_\mathrm{i}^{(c)}$ and $\pi-\Psi_\mathrm{a}^{(c)}\le \Psi \le \pi+\Psi_\mathrm{a}^{(c)}$, for the in-phase and anti-phase solutions, respectively.)
However, we expect (and confirm below) that the synchrony domain is smaller than for the two-oscillator case due to cluster instability.

\subsubsection{Stability of the two-cluster solution in the second-order approximation}
We quantify the transversal stability of the $(1+1)$ state using Eqs.~(\ref{TLE_general_form_a},\ref{TLE_general_form_b}). Substituting the mean fields for the $(1+1)$
state $Z^{(1)}=\ee^{\ii\Theta}$, $Z^{(2)}=\ee^{\ii\Phi}$, and $Z_2^{(1)}=\ee^{2\ii\Theta}$, $Z_2^{(2)}=\ee^{2\ii\Phi}$, we obtain 
\begin{eqnarray}
\Lambda_1( \Psi; \alpha, \e, \kp ) &=& - \frac{\e}{2} \biggl(\cos \alpha + \cos(\Psi-\alpha) - \frac{\e}{4\kappa}
  \bigl[ 1 - 2\cos(\Psi-2\alpha) + \cos(2\Psi) + 2\cos\Psi- 2\cos(2\Psi-2\alpha) \bigr]\biggr)\,,
  \label{eqn:Lambda1}\\
\Lambda_2( \Psi; \alpha, \e, \kp ) &=& - \frac{\e}{2} \biggl(\cos \alpha + \cos(\Psi+\alpha) - \frac{\e}{4\kappa} 
  \bigl[ 1 - 2\cos(\Psi+2\alpha) + \cos(2\Psi) + 2\cos\Psi - 2\cos(2\Psi+2\alpha)\bigr]\biggl)\,.
 \label{eqn:Lambda2}
 \end{eqnarray}
Thus, clusters are stable if the expressions in the brackets are (on average) positive.

Below, we discuss separately synchronous and asynchronous cluster states. Synchrony in this context means that two clusters adjust their frequencies and have some constant phase shift. More precisely, the synchronous cluster state
corresponds to a stable fixed point $\Psi_*$
of the reduced Eq.~(\ref{Adler_equation_1+1}),
while the asynchronous cluster state
corresponds to its drifting solution.
Note that according to our previous consideration
of Adler equation~(\ref{Adler}), there are two types
of synchronous cluster states:
synchronous in-phase states (with $\cos\Psi_* > 0$)
and synchronous anti-phase state (with $\cos\Psi_* < 0$).
Next, we want to analyze all these cluster states
for their transversal stability,
using the general framework developed at the beginning of this section.
Since the synchronous cluster state corresponds to a fixed point $\Psi_*$ of Eq.~(\ref{Adler_equation_1+1}),
the transversal Lyapunov exponents of this state are given by 
\begin{equation}
\lambda_{1,2} = \Lambda_{1,2}( \Psi_*; \alpha, \e, \kp )\,.
\label{eqn_TLE}
\end{equation}
On the other hand, in the case of an asynchronous
$(1+1)$ cluster state, Eq.~(\ref{Adler_equation_1+1})
has a drifting solution $\Psi(t)$.
Therefore, the expressions of $\Lambda_{1,2}(\cdot)$
depend on time, and the corresponding TLEs
are given by formulas
$$
\lambda_{1,2} = \lim\limits_{T\to\infty} \frac{1}{T} \int_0^T
\Lambda_{1,2}( \Psi(t); \alpha, \e, \kp ) \dd t\,,
$$
which we write in an equivalent form using the ergodicity property: 
\begin{equation}
\lambda_{1,2} = \frac{1}{C} \int_0^{2\pi} \frac{\Lambda_{1,2}( \psi; \alpha, \e, \kp )}{\left| \delta - \e\sin\psi \left(\cos\alpha - \frac{\e}{\kappa}\sin^2\alpha \cos\psi \right) \right|} \dd\psi\,,
  \label{eqn:lambda}
\end{equation}
where
\begin{equation}
  C = \int_0^{2\pi} \frac{\dd\psi}{\left| \delta - \e\sin\psi \left(\cos\alpha - \frac{\e}{\kappa}\sin^2\alpha \cos\psi \right) \right|}\,. 
  \label{eqn:C}
\end{equation}
Before presenting the results of TLE computation, we, for comparison, obtain TLE in the framework of the standard first-order phase reduction.

\subsubsection{Comparison of the results of the first- and second-order approximation}

We obtain the corresponding expressions neglecting the terms $\sim \e^2$ terms in Eqs.~(\ref{eqn:Lambda1},\ref{eqn:Lambda2},\ref{eqn:lambda},\ref{eqn:C}). They read
\begin{eqnarray}
\Lambda_{1,2}( \Psi, \alpha, \e, \kp ) &=& - \frac{\e}{2} \biggl(\cos \alpha + \cos(\Psi\mp \alpha)\biggr ) \,,
  \label{eqn:Lambda34}
 \end{eqnarray}
and 
\begin{equation}    
\lambda_{1,2} = \frac{1}{C} \int_0^{2\pi} \frac{\Lambda_{1,2}( \psi, \alpha, \e, \kp )}{\left| \delta - \e\sin\psi \cos\alpha  \right|} \dd\psi\,,\qquad
  C = \int_0^{2\pi} \frac{\dd\psi}{\left| \delta - \e\sin\psi \cos\alpha \right|}\,. 
  \label{lambda_running}
\end{equation}

\begin{figure}[!htb]
\centering
\includegraphics[width=0.46\columnwidth]{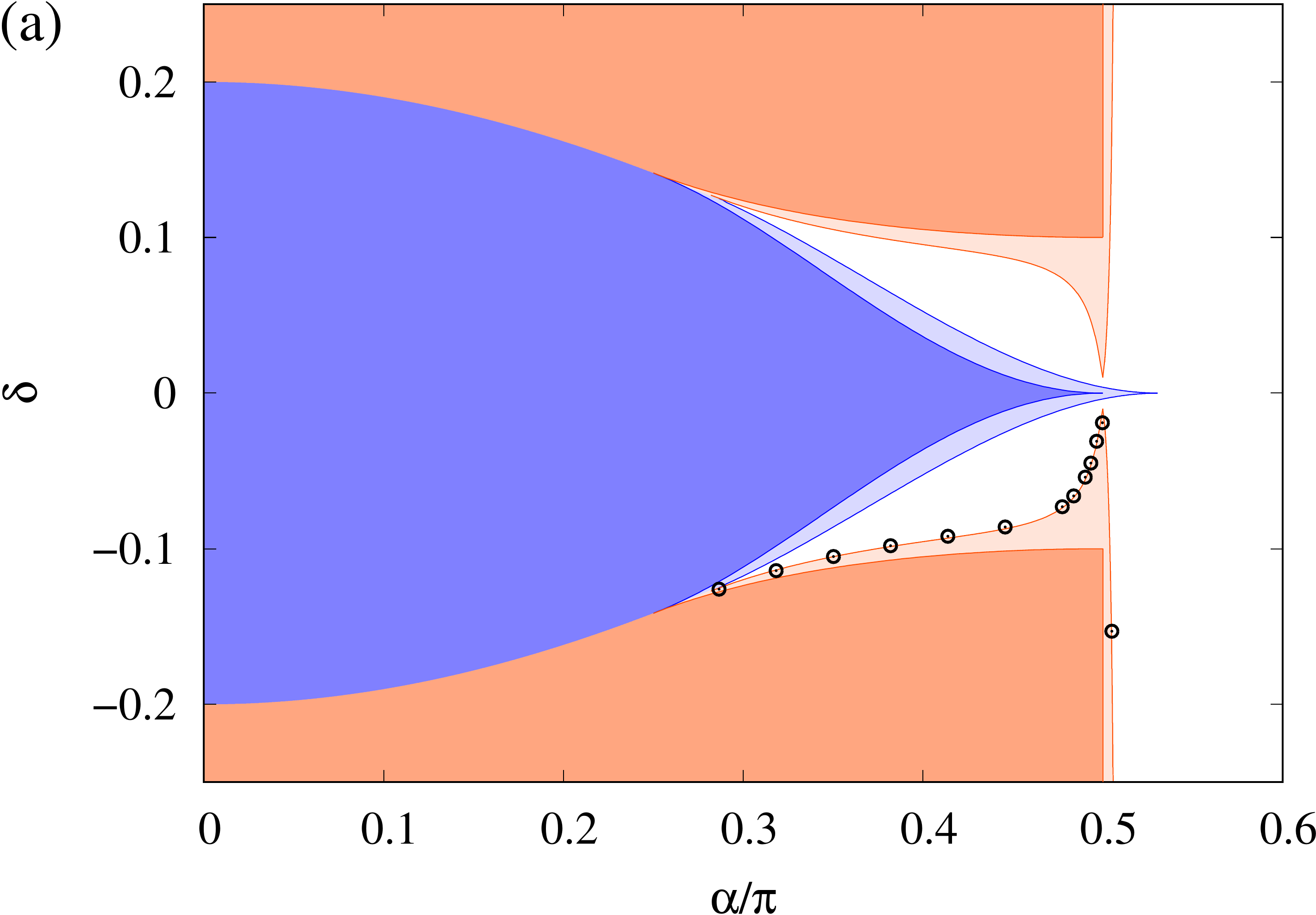}
\hspace{0.04\columnwidth}
\includegraphics[width=0.46\columnwidth]{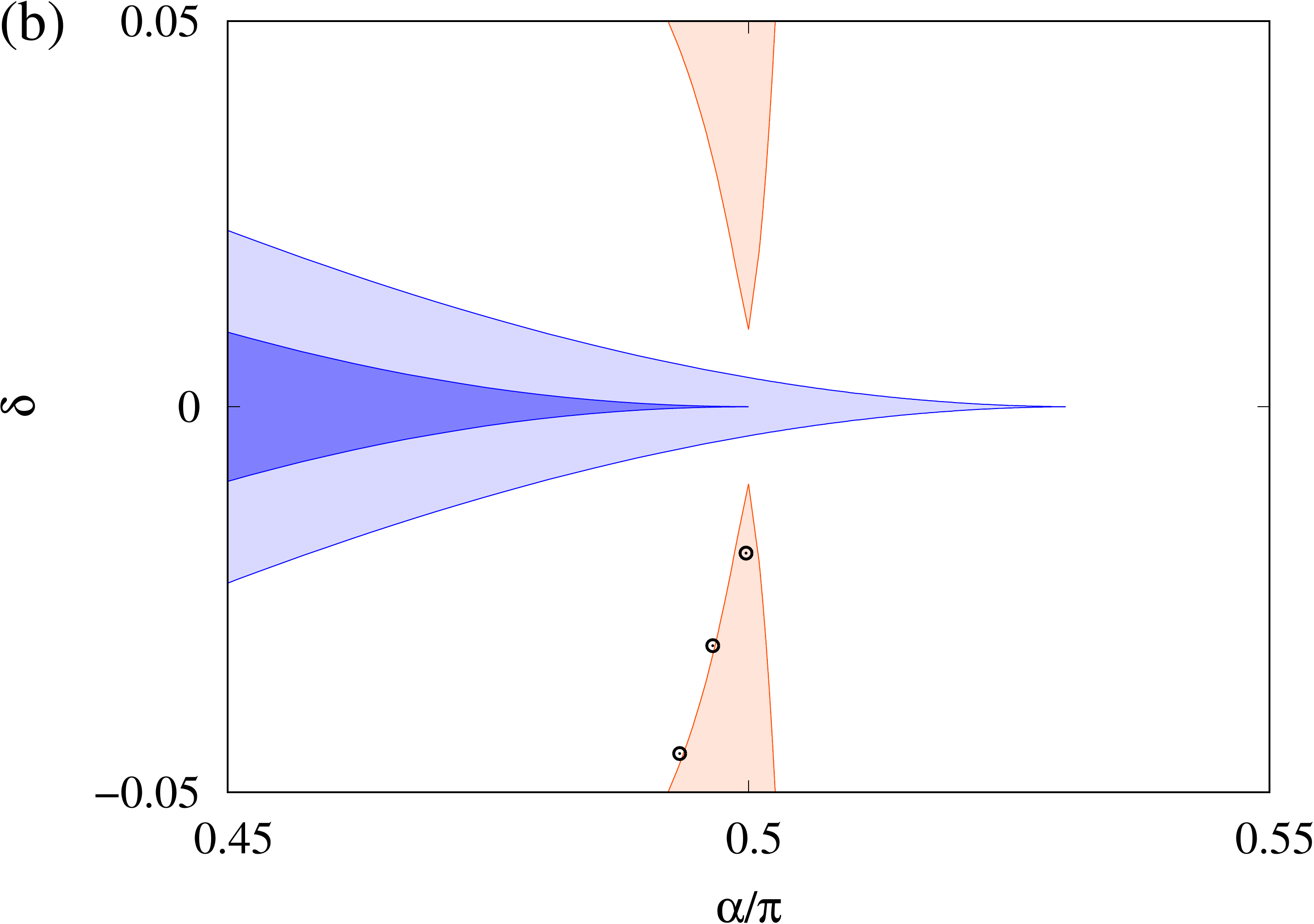}\\[2mm]
\includegraphics[width=0.46\columnwidth]{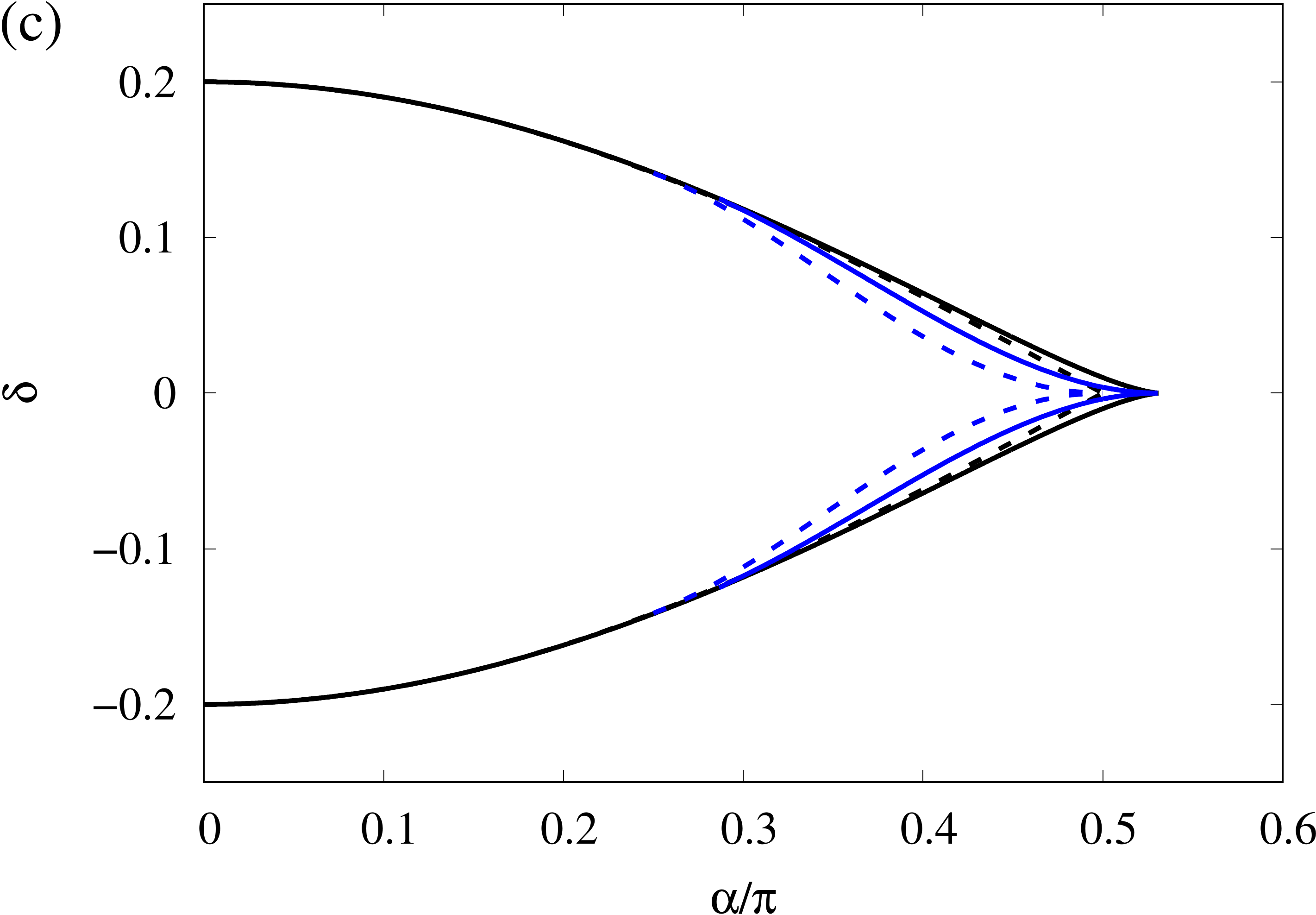}
\hspace{0.04\columnwidth}
\includegraphics[width=0.46\columnwidth]{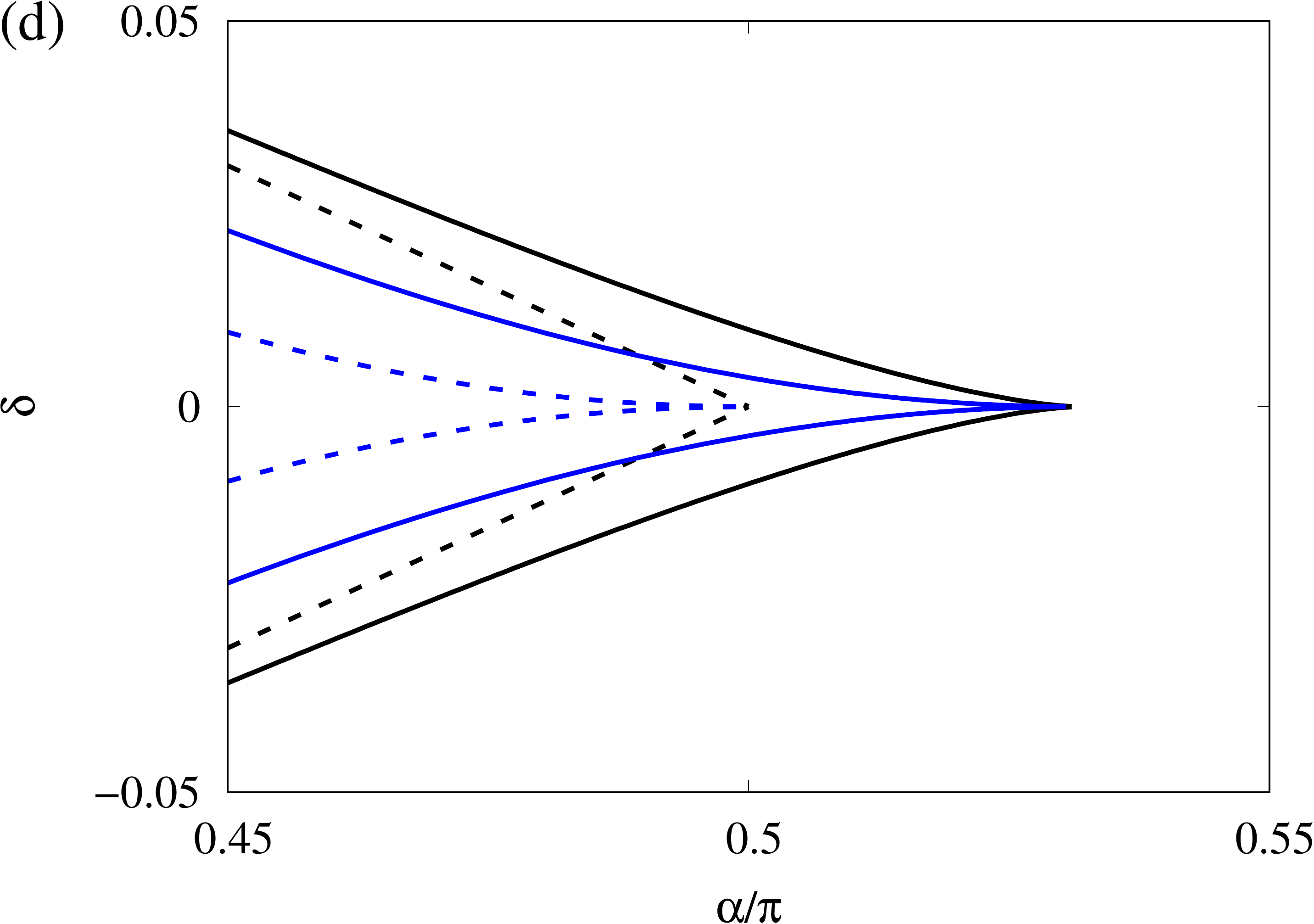}
\caption{
(a) Stability diagram for asynchronous
and synchronous in-phase cluster states
in model~(\ref{phase_eq_Leon}) with $\e = 0.2$.
Blue and orange shading indicate
synchronous and asynchronous cluster states.
Dark and light shading correspond to the results
of first- and second-order approximations, respectively.
Circles show stability boundaries
identified by dynamical continuation
in the Stuart-Landau oscillator model~(\ref{SL_2groups})
with $N_\mathrm{g} = 100$.
(c) Locking cone boundaries (black)
and transversal stability boundaries
of the synchronous in-phase cluster state (blue).
Dashed and solid lines show the results obtained from
the first- and second-order approximations, respectively.
Panels (b) and (d) are enlargements of the right parts of panels (a) and (c).
} 
\label{BifDiagram}
\end{figure}

\paragraph{Synchronous state.}
In the first-order approximation, we deal with the standard Adler equation 
\begin{equation}
\dot{\Psi} = \delta - \e \cos\alpha \sin\Psi\;.
\label{Eq:Adler:1}
\end{equation} 
If $\cos\alpha\ne 0$ its 
properties are determined by the ratio
\begin{equation}
\chi = \frac{\delta}{\e \cos\alpha}\;.
\label{Def:chi}
\end{equation}
If $|\chi|<1$, the Adler equation has two fixed points:
\begin{equation}
\Psi_\mathrm{i} = \arcsin \chi
\quad\mbox{and}\quad
\Psi_\mathrm{a} = \pi - \arcsin \chi\,.
\label{Adler:FP}
\end{equation} 
Moreover, if $\cos\alpha > 0$,
then $\Psi_\mathrm{i}$ is stable and $\Psi_\mathrm{a}$ is unstable.
On the contrary, if $\cos\alpha < 0$,
then $\Psi_\mathrm{i}$ is unstable and $\Psi_\mathrm{a}$ is stable (we remind that we assumed $\e>0$).
Recalling Eq.~(\ref{eqn_TLE}) and substituting $\Psi_\mathrm{i,a}$ into Eq.~(\ref{eqn:Lambda34}), we 
explicitly compute TLEs $\lambda_{1,2}^{(\mathrm{i})}$ and 
$\lambda_{1,2}^{(\mathrm{a})}$ for the in-phase and anti-phase locked states, respectively (see Appendix \ref{App1}). We demonstrate that the anti-phase solution is always unstable, while the in-phase locked state can be stable with respect to ``evaporation'' of oscillators from a cluster for certain values of $\alpha$ and $\delta$. Namely,
for $0<\alpha<\pi/4$, the clusters are stable within the whole locking domain determined by the Adler equation, $|\delta|<\e\cos\alpha$, while 
for $\pi/4 < \alpha < \pi/2$, the clusters lose their stability at the critical value 
$$
\delta_\mathrm{cr}^{(\mathrm{i})} = \pm \e \cos\alpha \sin 2\alpha\;.
$$
We illustrate these results in Fig.~\ref{BifDiagram}.

\paragraph{Asynchronous state.}

We recall that an asynchronous $(1+1)$ state considered in this section means the existence of two clusters that have different frequencies and are, therefore, not synchronized.
Indeed, if $|\chi| > 1$, then Eq.~(\ref{Eq:Adler:1}) has
a drifting solution, and we have to compute the integrals in Eq.~(\ref{lambda_running}). Performing these computations in Appendix~\ref{App1}, we find that the clusters are stable for $0 < \alpha < \pi/4$, while  for
$\pi/4 < \alpha < \pi/2$ they are stable if
$$
|\delta|>\delta_\mathrm{cr} = \frac{\e}{2 \sin\alpha}\,.
$$

In the second-order approximation, we find the boundaries of the
synchronization domain semi-analytically. 
More precisely, we obtain the stability 
boundaries of the synchronous in-phase state 
by solving numerically the equation
$$
\max\limits_{j=1,2} \Lambda_j(\Psi_*,\alpha,\e,\kappa) = 0,
$$
where $\Lambda_j(\Psi_*,\alpha,\e,\kappa)$
are given by the Eqs.~(\ref{eqn:Lambda1}), (\ref{eqn:Lambda2})
and the value of the fixed point $\Psi_*$
is determined from the relation
\begin{equation}
\delta = \e\cos\alpha\sin\Psi_* - \frac{\e^2}{4\kp} \biggl(1-\cos(2\alpha) \biggr)\sin(2\Psi_*)\;,
\label{Eq:FP:2}
\end{equation}
which holds for the fixed points of Eq.~(\ref{Adler_equation_1+1})
~\footnote{For the numerical solution, we exploit the Matlab function {\rm fsolve()}.}.
Moreover, using similar calculations
for the synchronous anti-phase state,
we found that this state is always transversally unstable.
Finally, the stability boundaries of the asynchronous state
were obtained from the condition
$\max\limits_{j=1,2} \lambda_j = 0$,
where $\lambda_j$ is found by solving numerically Eq.~(\ref{eqn:lambda}).
The results of the above numerical-analytical
consideration are summarized in the stability diagram
shown in Fig.~\ref{BifDiagram}.

\begin{figure}[!htb]
\centering
\includegraphics[width=\columnwidth]{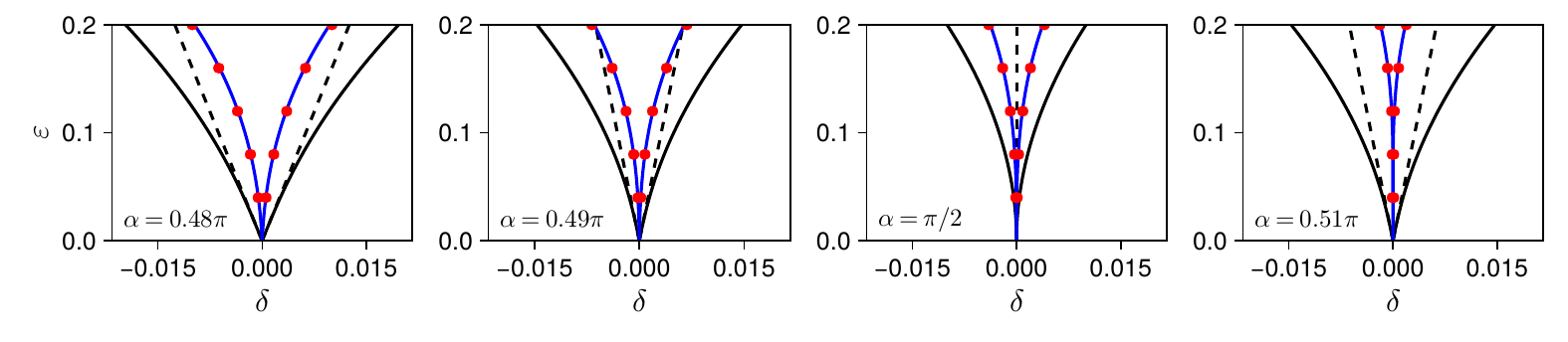}
\caption{Full synchronization tongues in the $(1+1)$ states for different values of $\alpha$. 
Black curves correspond to the two-oscillator tongues shown in Fig.~\ref{Tongues}. Dashed black curves are the 
first-order tongues, solid-black curves are the second-order tongues. 
The blue curves correspond to the $(1+1)$ cluster stability boundaries, obtained 
from Eqs.~(\ref{eqn:Lambda1},~\ref{eqn:Lambda2}). Red circles show the stability boundaries identified by
dynamical continuation in the full Stuart-Landau system with $N_g=100$.
} 
\label{FS_boundary_1+1}
\end{figure}

Figure~\ref{FS_boundary_1+1} illustrates further the cross-section of stability domains in the $(\delta,\e)$ plane, showing the Arnold tongues. 
Here, for four values of $\alpha$ in the vicinity of $\pi/2$, we compare 
the locking tongues of the genuine two-oscillator system (black curves) with the domain in which the $(1+1)$ cluster state is both locked and transversally stable (blue curves). The latter domain is the 
full synchronization tongue, because only inside it do all $2N_g$ units remain grouped into two mutually locked clusters.
Close to the boundary of this domain,
the phase shift $\Psi$ approaches its maximal value 
$\Psi_{\mathrm{i},\mathrm{a}}$, and the clusters evaporate before the locking is lost.
The red circles, obtained by dynamical continuation in the full Stuart-Landau system~(\ref{SL_2groups}) with 
$N_g=100$, closely follow the analytically computed boundaries, which again demonstrates the accuracy of the second-order reduction.

\subsection{Three clusters}

Another cluster state that we observe is the $(2+1)$ state, when the group with higher frequency 
$\omega_1>\omega_2$ splits into two equally sized clusters with phases $\Theta_1$ and $\Theta_2$, 
while the oscillators in the second group merge into a single cluster with phase $\Phi_1$.
The dynamics of this state is determined by Eqs.~(\ref{cluster_eq_generic_a}) and (\ref{cluster_eq_generic_b}), 
where we set $p=2$ and $q=1$:
\begin{eqnarray}
    \dot{\Theta}_1 &=& \Omega_1 + \mathrm{Im}\left[f_1\left(Z^{(1,2)},Z_2^{(1,2)}; \e, \alpha, \kappa\right)\cdot\ee^{-\ii\Theta_1} 
    + f_2\left(Z^{(1,2)},Z_2^{(1,2)}; \e, \alpha, \kappa\right)\cdot\ee^{-2\ii\Theta_1} \right]\,,\label{eqn:2+1_a}\\
    \dot{\Theta}_2 &=& \Omega_1 + \mathrm{Im}\left[f_1\left(Z^{(1,2)},Z_2^{(1,2)}; \e, \alpha, \kappa\right)\cdot\ee^{-\ii\Theta_2} 
    + f_2\left(Z^{(1,2)},Z_2^{(1,2)}; \e, \alpha, \kappa\right)\cdot\ee^{-2\ii\Theta_2} \right]\,,\label{eqn:2+1_b}\\
    \dot{\Phi}_1 &=& \Omega_2 + \mathrm{Im}\left[f_1\left(Z^{(1,2)},Z_2^{(1,2)}; \e, \alpha, \kappa\right)\cdot\ee^{-\ii\Phi_1} 
    + f_2\left(Z^{(1,2)},Z_2^{(1,2)}; \e, \alpha, \kappa\right)\cdot\ee^{-2\ii\Phi_1} \right]\,,\label{eqn:2+1_c}
\end{eqnarray}
and where $Z^{(1)}=(\ee^{\ii\Theta_1}+\ee^{\ii\Theta_2})/2$ and $Z^{(2)}=\ee^{\ii\Phi_1}$.
The stability of this state depends on two factors.
First, the configuration $\Theta_1(t)$, $\Theta_2(t)$, and $\Phi_1(t)$ should be a stable solution of the system~\eqref{eqn:2+1_a}--\eqref{eqn:2+1_c}.
We call this macroscopic stability.
Second, each of  the three clusters should be transversally stable
such that neither cluster evaporates.
This type of stability can be analyzed
by the instantaneous growth-rate factors
given by Eqs.~(\ref{TLE_general_form_a}) 
and (\ref{TLE_general_form_b}).
For the specific $(2+1)$ state, they read
\begin{eqnarray}
    \Lambda_{\Theta_1} &=& 
    -\mathrm{Re}\left[f_1\left(Z^{(1,2)},Z_2^{(1,2)};\e,\alpha,\kappa\right)\cdot\ee^{-\ii \Theta_1}
    +2f_2\left(Z^{(1,2)},Z_2^{(1,2)};\e,\alpha,\kappa\right)\cdot\ee^{-2\ii \Theta_1}\right], 
    \label{TLE_2+1_1a} \\
    \Lambda_{\Theta_2} &=& 
    -\mathrm{Re}\left[f_1\left(Z^{(1,2)},Z_2^{(1,2)};\e,\alpha,\kappa\right)\cdot\ee^{-\ii \Theta_2}
    +2f_2\left(Z^{(1,2)},Z_2^{(1,2)};\e,\alpha,\kappa\right)\cdot\ee^{-2\ii \Theta_2}\right], 
    \label{TLE_2+1_1b} \\
    \Lambda_{\Phi_1} &=& 
    -\mathrm{Re}\left[f_1\left(Z^{(1,2)},Z_2^{(1,2)};\e,\alpha,\kappa\right)\cdot\ee^{-\ii \Phi_1}
    +2f_2\left(Z^{(1,2)},Z_2^{(1,2)};\e,\alpha,\kappa\right)\cdot\ee^{-2\ii \Phi_1}\right],
    \label{TLE_2+1_2}
\end{eqnarray}
where the factors $\Lambda_{\Theta_1}$, $\Lambda_{\Theta_2}$, and $\Lambda_{\Phi_1}$,
correspond to clusters with phases $\Theta_1$, $\Theta_2$, and $\Phi_1$, respectively.

\begin{figure}[!htb]
\centering
\includegraphics[height=0.23\textwidth]{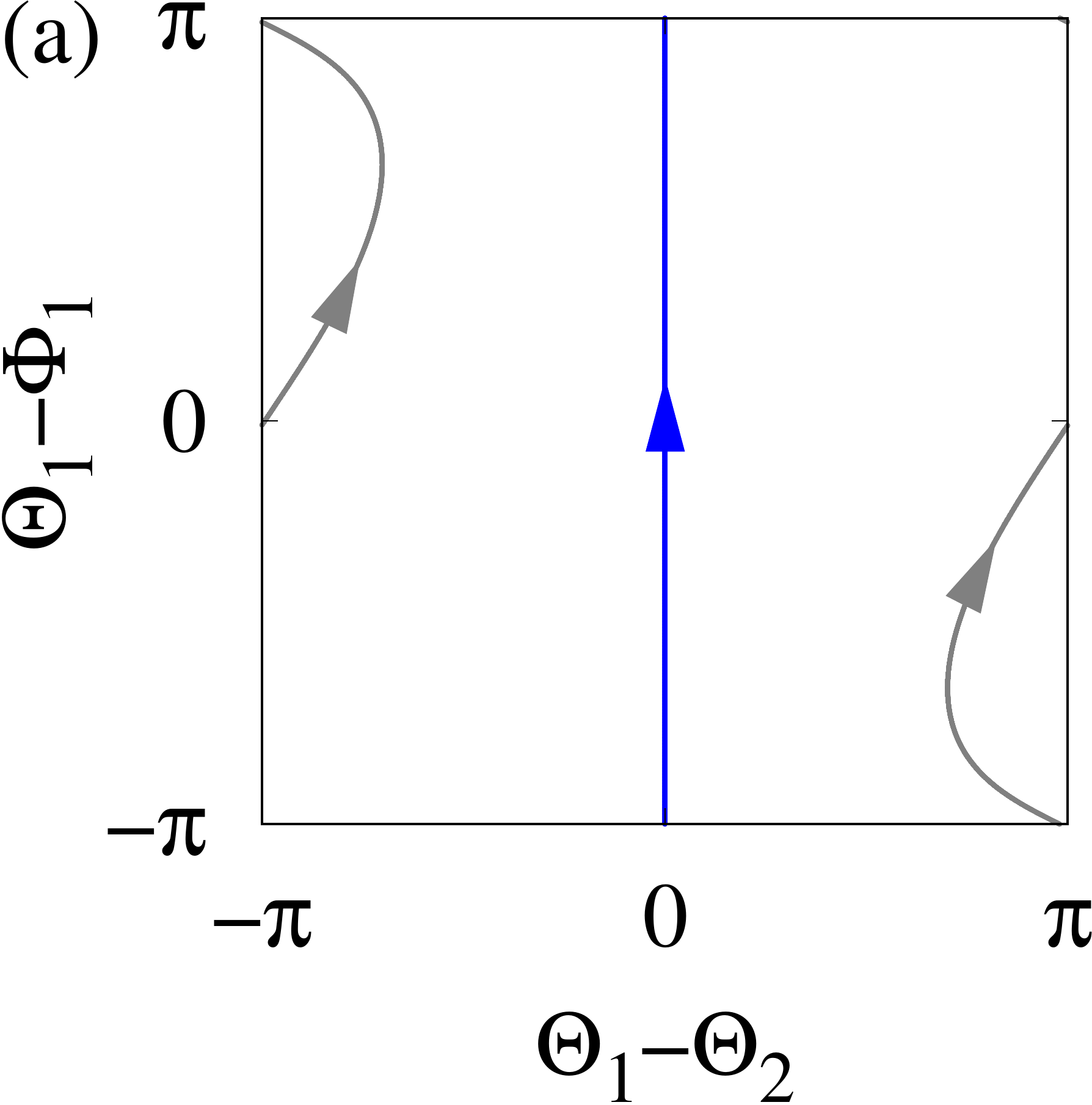}\hspace{0.03\textwidth}%
\includegraphics[height=0.23\textwidth]{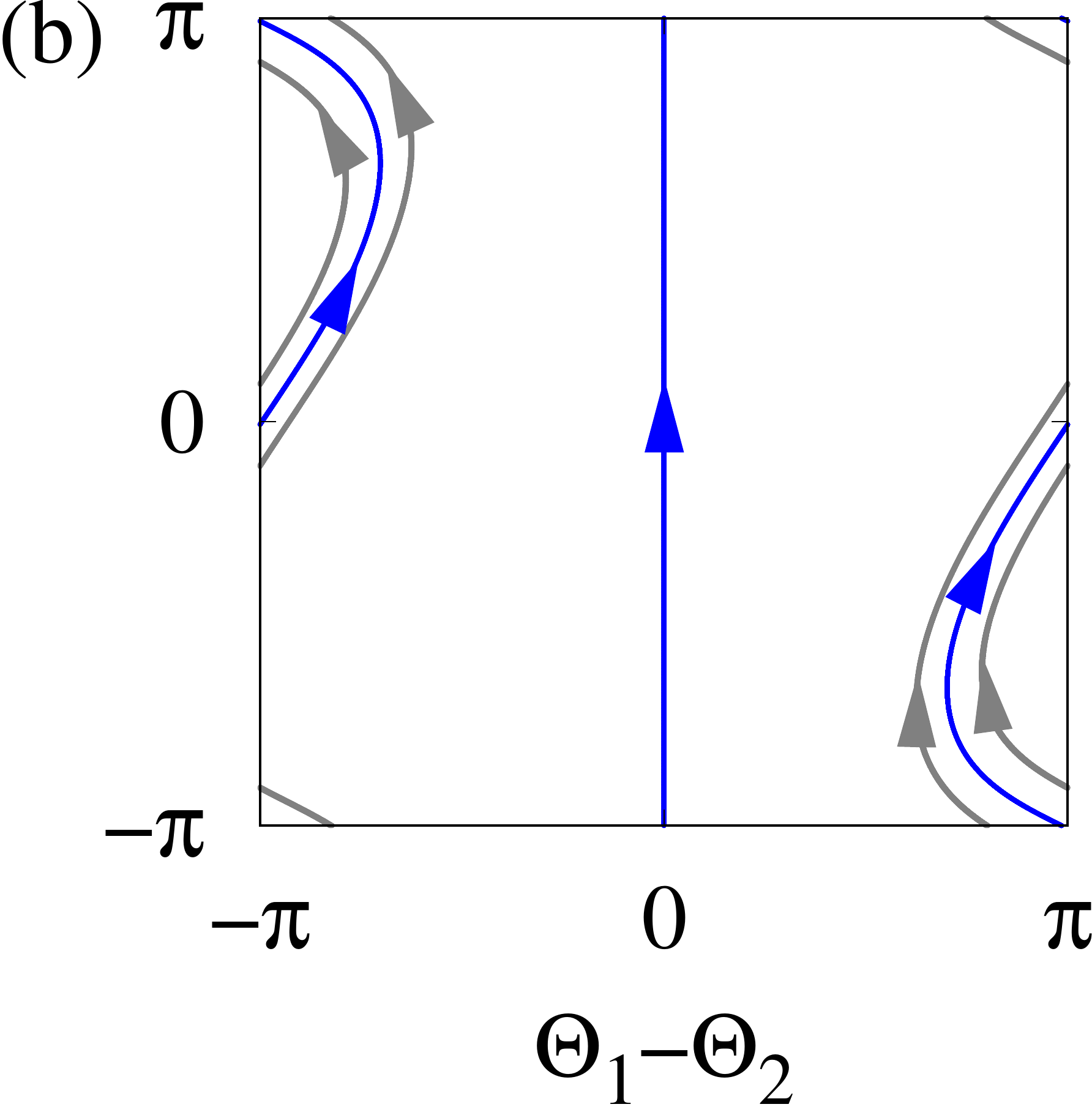}\hspace{0.03\textwidth}%
\includegraphics[height=0.23\textwidth]{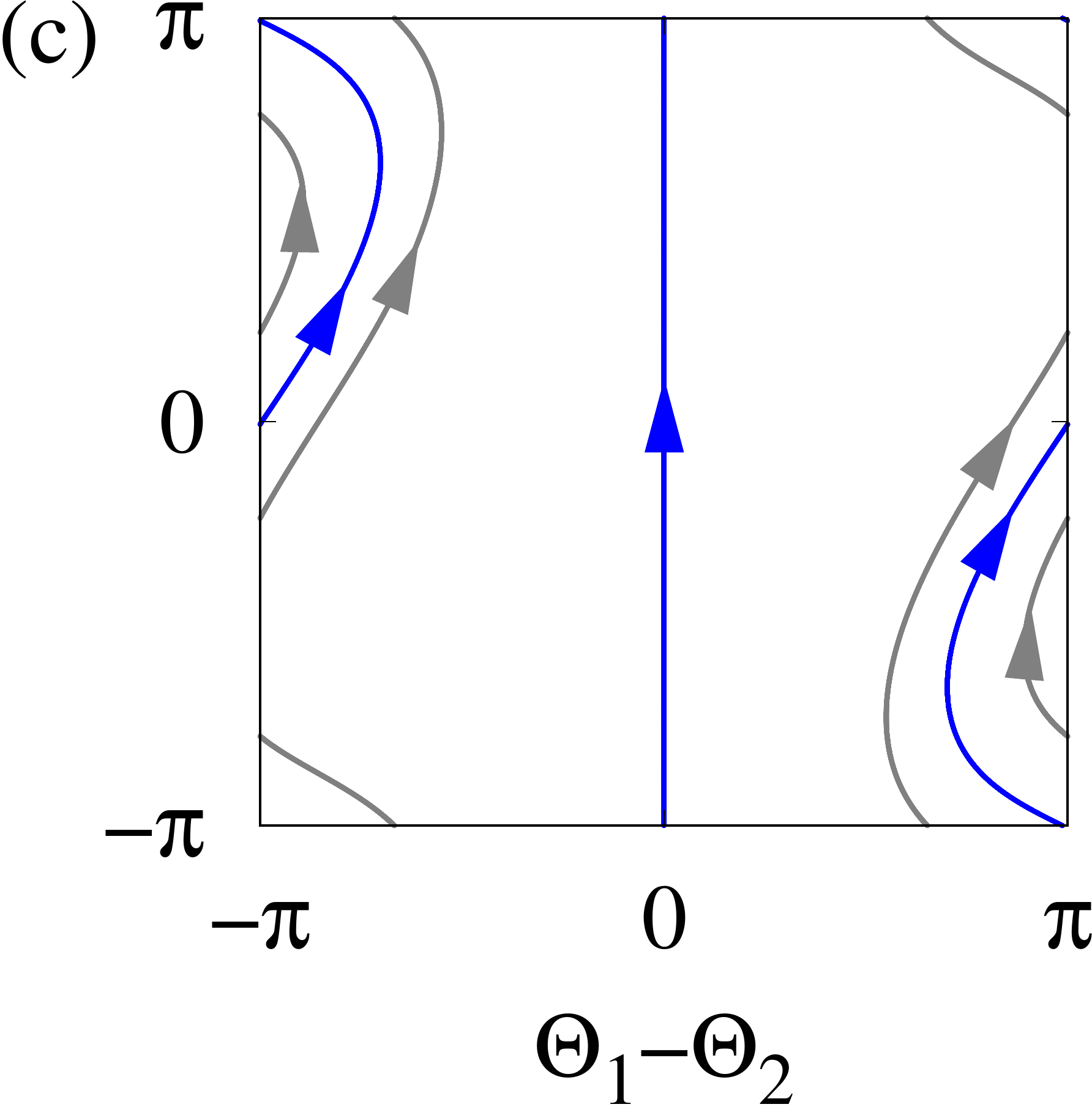}\hspace{0.03\textwidth}%
\includegraphics[height=0.23\textwidth]{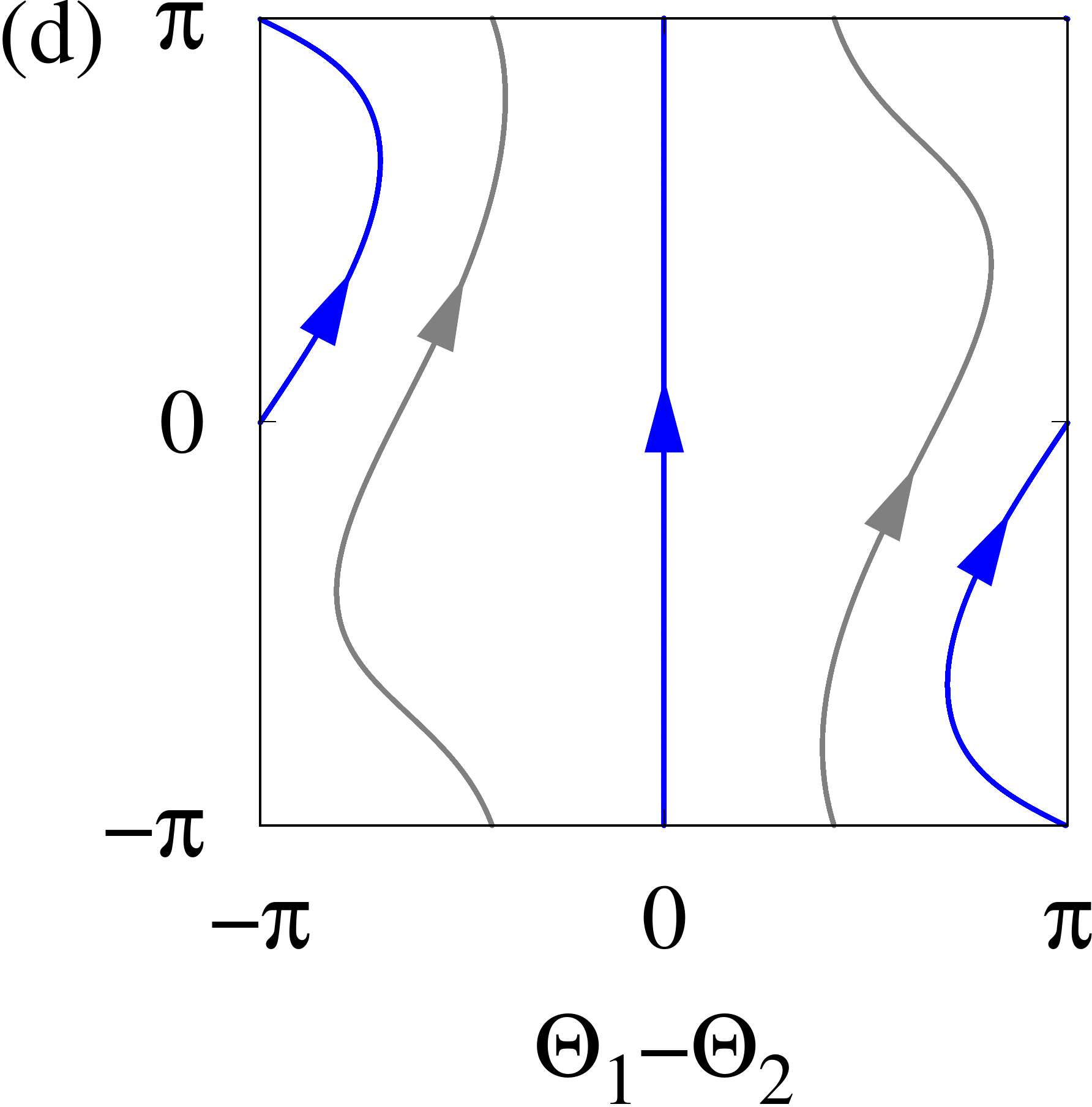}
\caption{Subcritical pitchfork bifurcation
of limit cycles in the three-group model
corresponding to the $(2+1)$ state.
Blue/dark curves and grey/light curves
show stable and unstable limit cycles, respectively.
(a) $\alpha = 0.46\pi$, (b) $\alpha = 0.468\pi$,
(c) $\alpha = 0.47\pi$, and (d) $\alpha = 0.48\pi$.
The oscillators $\Theta_1$ and $\Theta_2$
have frequency $\omega_1 = 2$,
while the oscillator $\Phi_1$
has frequency $\omega_2 = 1.7$.
Other parameters: $\e = 0.2$.
} 
\label{Pitchfork}
\end{figure}

System~\eqref{eqn:2+1_a}--\eqref{eqn:2+1_c} allows for a detailed bifurcation analysis,
since using the nondegenerate linear transformation
$(\Theta_1,\Theta_2,\Phi_1)\rightarrow ((\Theta_1+\Theta_2)/2, (\Theta_1-\Theta_2)/2, \Theta_1-\Phi_1)$, it can be rewritten in an equivalent form with a r.h.s.,
which depends only on two phase differences $\Theta_1-\Theta_2$ and $\Theta_1-\Phi_1$
(see Appendix~\ref{App2}). This means that the dynamics of system~\eqref{eqn:2+1_a}--\eqref{eqn:2+1_c} is mainly determined by the dynamics
of the corresponding two-dimensional system
on a torus parameterized by $\Theta_1-\Theta_2$ and $\Theta_1-\Phi_1$.
By fixing the frequencies $\omega_1>\omega_2$ in this system
and varying the phase lag $\alpha$ from small to large values,
we typically observe the following bifurcation scenario,
see Fig.~\ref{Pitchfork}. For small values of $\alpha$,
system~\eqref{eqn:2+1_a}--\eqref{eqn:2+1_c} has two solutions:
a stable symmetric solution with $\Theta_1 = \Theta_2$
and an unstable asymmetric solution with $\Theta_1 \ne \Theta_2$.
Note that only the latter solution corresponds to a $(2+1)$ cluster state,
while the former solution represents a $(1+1)$ cluster state.
Therefore, $(2+1)$ cluster states are macroscopically unstable for small $\alpha$.
As $\alpha$ increases, the stability of the symmetric solution remains unchanged,
while the unstable solution with $\Theta_1 \ne \Theta_2$
undergoes a subcritical pitchfork bifurcation of limit cycles,
leading to the appearance of one stable and two unstable asymmetric solutions.
The occurrence of such a bifurcation is not surprising,
since system~\eqref{eqn:2+1_a}--\eqref{eqn:2+1_c} is reflection symmetric
with respect to the permutation of $\Theta_1$ and $\Theta_2$.
We emphasize that only after this bifurcation,
can stable $(2+1)$ cluster state be expected
(although its transversal stability still needs to be verified).
Importantly, unstable asymmetric solutions arising
from the subcritical pitchfork bifurcation
act as boundaries in the phase space, separating the
basins of attraction for stable symmetric and asymmetric solutions.
As a result, bistability between the $(2+1)$ cluster state
and the $(1+1)$ cluster state becomes possible.

\begin{figure}[!htb]
\centering
\includegraphics[width=0.5\columnwidth]{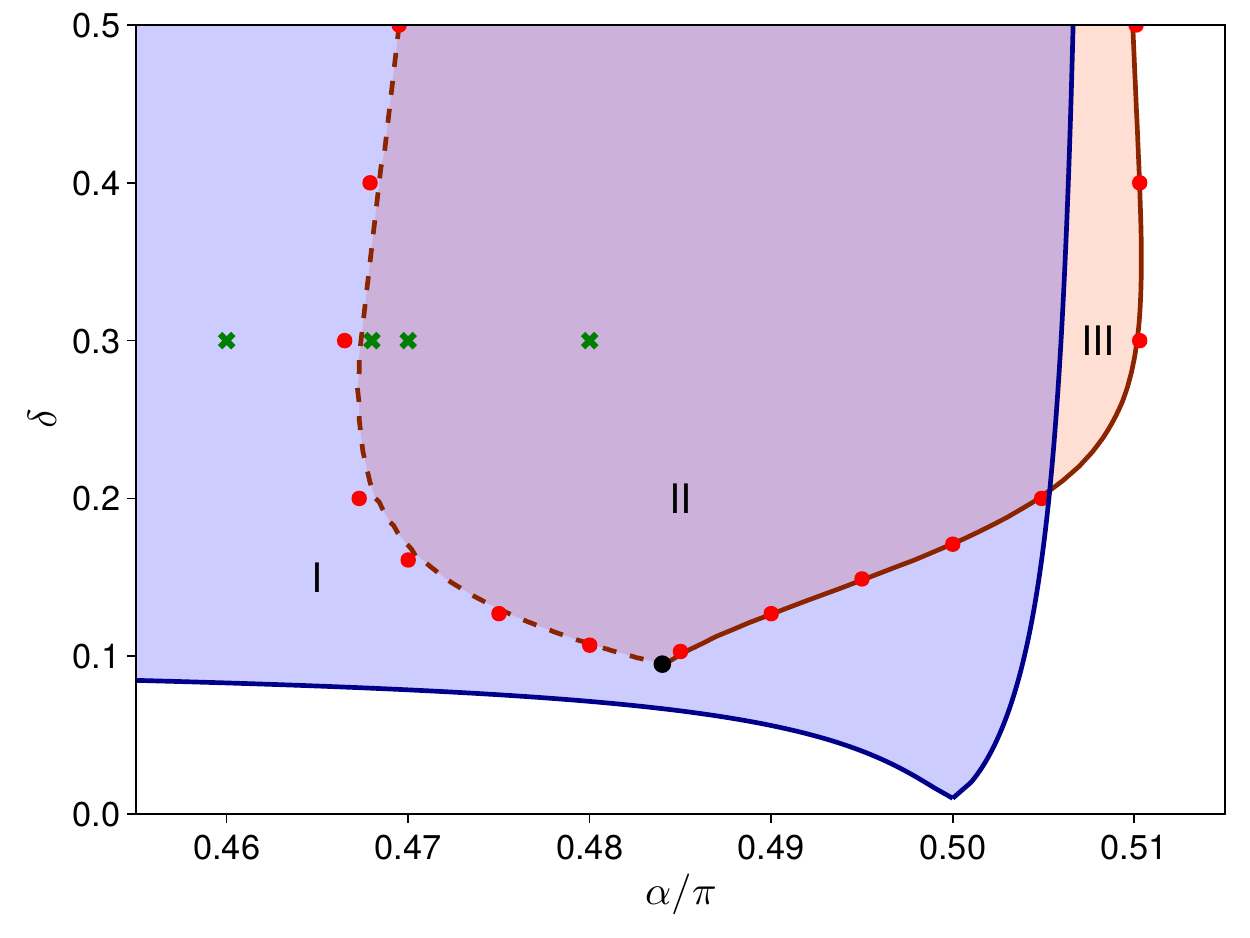}
\caption{Stability diagram for the three-cluster $(2+1)$ and two-cluster $(1+1)$ states for $\e=0.2$. 
The stability boundaries of the two-cluster state are shown by blue curves, and the boundaries of existence of the three-cluster states are given by the solid and dashed orange curves, to indicate two different mechanisms of three-cluster state destruction (see text). 
Red circles show the stability boundaries identified by simulating the Stuart-Landau system 
with $N_g=100$ oscillators.
In region I, we observe only the $(1+1)$ states, in region II, we observe bistability, 
when both $(1+1)$ and $(2+1)$ states coexist, and in region III, we observe only the $(2+1)$ states.
The parameter values used to plot four panels in Fig.~\ref{Pitchfork} are marked with green crosses.
} 
\label{Domain_2+1}
\end{figure}

To find the parameter values for which the subcritical pitchfork bifurcation in 
system~\eqref{eqn:2+1_a}--\eqref{eqn:2+1_c} takes place, we used the following numerical scheme. First,
for a given set of parameters $(\alpha,\delta)$ where the asymmetric solution is stable, we initialized 
the system within the basin of attraction of the $(2+1)$ state. Next, we simulated the system for 
$5\cdot 10^5$ time units to skip the transient, and then calculated the time-average $R^{(1)}$ of the
modulus $|Z^{(1)}(t)|=|\ee^{\ii\Theta_1}+\ee^{\ii\Theta_2}|/2$ over the 
next $10^4$ time units. The value of $R^{(1)}$ served as an indicator of the pitchfork bifurcation. 
If $R^{(1)}<1$, we incrementally changed one parameter while holding the other constant, 
using the final state of the system as the initial condition for the next step. This continuation was 
performed independently in two directions: varying $\delta$ in steps of $0.0001$ for a fixed $\alpha$, 
and varying $\alpha$ in steps of $0.0001\pi$ for a fixed $\delta$. At a critical point 
$(\alpha_c,\delta_c)$ where the pitchfork bifurcation takes place, the value of $R^{(1)}$ jumps abruptly
to $1$. At this point, the asymmetric solution with $R^{(1)}<1$ becomes unstable and the symmetric
solution with $R^{(1)}=1$ remains the only attractor. The results obtained using this algorithm are shown 
by the dashed curve in Fig.~\ref{Domain_2+1}.

In the case when system~(\ref{eqn:2+1_a})-(\ref{eqn:2+1_c}) has a stable asymmetric solution,
we calculated the TLEs of the corresponding
$(2+1)$ cluster state.
For this, we solved system~(\ref{eqn:2+1_a})-(\ref{eqn:2+1_c}) numerically, inserted the obtained values $\Theta_1(t)$, $\Theta_2(t)$, $\Phi_1(t)$
into formulas~\eqref{TLE_2+1_1a}--\eqref{TLE_2+1_2}
and finally performed the time averaging according to formula~\eqref{averageTLE} to obtain three TLEs.
Detecting the parameter values
for which one of the TLEs becomes positive,
we find the transversal stability 
boundary of the $(2+1)$ cluster state.
The numerical results show that, typically,
the slowest cluster is the first to lose its stability.
This instability occurs at the points
on the solid red curve in Fig.~\ref{Domain_2+1}.
Together with the bifurcation analysis
of system~(\ref{eqn:2+1_a})-(\ref{eqn:2+1_c}),
this implies that $(2+1)$ cluster states
are stable (both macroscopically and transversally)
in the union of the shaded regions II and III
in Fig.~\ref{Domain_2+1}.
For completeness, we also show on the same diagram
the previously calculated stability region
of the asynchronous cluster state
from Fig.~\ref{BifDiagram}.
According to this diagram, the $(1+1)$ and $(2+1)$ cluster states can coexist stably, as in region II,
or only one of these states can be stable,
as in regions I and III.
Finally, we notice that the stability boundaries
of the $(2+1)$ cluster state determined
using the second-order phase reduced model
are in good agreement with the stability boundaries
obtained from the direct numerical simulations
in the full Stuart-Landau system with $N_g=100$ oscillators (red dots in Fig.~\ref{Domain_2+1}).

\section{Discussion}

In this paper, we explored the benefit of the second-order phase approximation for describing the dynamics of networks of Stuart-Landau oscillators. The object of our analysis was a two-group population with global coupling, quantified by the complex coupling coefficient $\e\ee^{\ii\alpha}$.

As is well known, the first-order phase approximation for coupled Stuart-Landau units yields the Kuramoto model with terms $\sim\e$. The second-order model additionally contains the terms $\sim\e^2/\kappa$, with the Floquet exponent $\kappa$, which are essential for strong coupling and/or weak limit cycles' stability. Therefore, in many cases, the second-order model provides only a quantitative improvement to the first-order description. However, there are also examples where the inclusion of second-order terms yields qualitative effects \citep{kumar2021,Mau-Omelchenko-Rosenblum-24,bick2024a}.
Our study provides further examples of this kind. More specifically, the second-order terms turn out to be crucial for describing the dynamics of our system for $\alpha \approx \pi/2$ (nearly neutral coupling). 
For example, for two coupled oscillators, only the second-order terms describe the bistability between in-phase and anti-phase locking. At the same time, in the first-order approximation, it does not appear. Furthermore, using the second-order model, we can describe three-cluster states in which one group of identical Stuart-Landau units splits into two halves, while the other forms a single cluster. Although, according to the Watanabe-Strogatz theory \citep{Watanabe-Strogatz-93,Watanabe-Strogatz-94,Pikovsky-Rosenblum-15}, such configurations are impossible in the first-order model.

We also want to emphasize another aspect of our results.
The Kuramoto model has a relatively simple mathematical form, which makes it easier to consider analytically. On the other hand, the second-order phase reduction leads to much more complicated equations, and a natural question arises
as to whether they have any practical utility. 
Several issues follow immediately from the structure of these equations. So, triplet terms in the second-order model reveal the emergence of a phase-oscillator hypernetwork in pairwise-coupled Stuart-Landau oscillators; the appearance of pairwise coupling terms for non-connected units explains the remote synchrony \citep{kumar2021}. Moreover, the dependence of the coupling terms on the frequencies of oscillatory units explains the 
observation of the frequency dependence in the process of coupling function reconstruction from observations \citep{Blaha_et_al-11}.
More remarkable is that our second-order model admits analytical investigation. In particular, our results regarding the behavior of two interacting units are analytical. In addition, we recall that the domains of existence of the $(1+1)$ and $(2+1)$ cluster states were found semi-analytically (by numerically evaluating some explicitly defined integrals). 
Overall, this demonstrates that second-order phase reduction is not an abstract mathematical construct, but a useful tool for analyzing oscillatory networks.

As a final remark, we note that third- and higher-order models can, in principle, be constructed \citep{mau2023c}. However, analytical treatment of such models is hardly possible, and using them for simulation offers no advantage, since simulating the original system is typically faster.

\acknowledgments

The work of O.E.O. was supported
by the Deutsche Forschungsgemeinschaft under Grant
No. OM 99/2-3.

\section*{Author declarations}

{\bf Conflict of Interest}

The authors have no conflicts to disclose.

\section*{Author contributions}

{\bf Y.~Baibolatov:} Conceptualization (equal); Investigation (equal); Methodology (equal); Validation (equal); Writing -- original draft (equal); Writing -- review \& editing (equal).
{\bf O.~E.~Omel'chenko:} Conceptualization (equal); Investigation (equal); Methodology (equal); Validation (equal); Writing -- original draft (equal); Writing -- review \& editing (equal).
{\bf M.~Rosenblum:} Conceptualization (equal); Investigation (equal); Methodology (equal); Validation (equal); Writing -- original draft (equal); Writing -- review \& editing (equal).

\section*{Data availability}

Data sharing is not applicable to this article as no new data were created or analyzed in this study.

\appendix
\section{Computing the transverse Lyapunov exponents in the first-order approximation}
\label{App1}

In the first-order approximation,
the expressions of the TLEs of the $(1+1)$ cluster state are relatively simple.
They can be written explicitly
for both synchronous and asynchronous states.
The resulting formulas allow for further investigation,
so that in this case the transversal instability boundaries (red curves in Fig.~\ref{BifDiagram})
can also be described analytically.

\subsection{Synchronous state}

Substituting $\Psi_\mathrm{i,a}$ given by formulas~\eqref{Adler:FP}
into Eq.~(\ref{eqn:Lambda34}), we obtain:
\begin{eqnarray*}
\lambda_1^{(\mathrm{i})} &=& - \frac{\e}{2} \left( \cos\alpha + \cos(\Psi_\mathrm{i} -\alpha)\right)
= - \frac{\e}{2} \left( \cos\alpha + \cos\Psi_\mathrm{i} \cos\alpha + \sin\Psi_\mathrm{i} \sin\alpha \right) \\[2mm]
&=& - \frac{\e}{2} \left( ( 1 + \sqrt{1 - \chi^2} ) \cos\alpha + \chi \sin\alpha \right),\\[2mm]
\lambda_2^{(\mathrm{i})} &=& - \frac{\e}{2} \left( \cos\alpha + \cos(\Psi_\mathrm{i} + \alpha)\right)
= - \frac{\e}{2} \left( \cos\alpha + \cos\Psi_\mathrm{i} \cos\alpha - \sin\Psi_\mathrm{i} \sin\alpha \right) \\[2mm]
&=& - \frac{\e}{2} \left( ( 1 + \sqrt{1 - \chi^2} ) \cos\alpha - \chi \sin\alpha \right),
\end{eqnarray*}
and
\begin{eqnarray*}
\lambda_1^{(\mathrm{a})} &=& - \frac{\e}{2} \left( \cos\alpha + \cos(\Psi_\mathrm{a} -\alpha)\right)
= - \frac{\e}{2} \left( ( 1 - \sqrt{1 - \chi^2} ) \cos\alpha + \chi \sin\alpha \right),\\[2mm]
\lambda_2^{(\mathrm{a})} &=& - \frac{\e}{2} \left( \cos\alpha + \cos(\Psi_\mathrm{a} + \alpha)\right)
= - \frac{\e}{2} \left( ( 1 - \sqrt{1 - \chi^2} ) \cos\alpha - \chi \sin\alpha \right),
\end{eqnarray*}
where $\chi = \delta/(\varepsilon \cos\alpha)$.

Using these expressions of $\lambda_1^{(\mathrm{a})}$ and $\lambda_2^{(\mathrm{a})}$,
it can be shown that synchronous anti-phase cluster states
are always unstable for $\chi\ne 0$.
For this, we recall that such states correspond
to stable fixed points of Eq.~(\ref{Eq:Adler:1})
only if $\cos\alpha < 0$,
i.e. for $\pi/2 < |\alpha| < \pi$.
In addition, we note that
\begin{eqnarray*}
\lambda_1^{(\mathrm{a})} &=& - \frac{\e}{2} ( 1 - \sqrt{1 - \chi^2} ) \cos\alpha\:
\left(
1 + \frac{\chi}{ 1 - \sqrt{1 - \chi^2}} \tan\alpha
\right),\\[2mm]
\lambda_2^{(\mathrm{a})} &=& - \frac{\e}{2} ( 1 - \sqrt{1 - \chi^2} ) \cos\alpha\:
\left(
1 - \frac{\chi}{ 1 - \sqrt{1 - \chi^2}} \tan\alpha
\right).
\end{eqnarray*}
Now it is clear that if $\chi\tan\alpha > 0$ and $\cos\alpha < 0$, then $\lambda_1^{(\mathrm{a})} > 0$.
On the other hand, if $\chi\tan\alpha < 0$ and $\cos\alpha < 0$, then $\lambda_2^{(\mathrm{a})} > 0$.
This means unconditional transversal instability
of synchronous anti-phase cluster states.

Unlike synchronous anti-phase cluster states,
their in-phase counterparts can be transversely stable
for certain values of $\alpha$ and $\delta$.
To find these values, we consider the equations
$\lambda_1^{(\mathrm{i})} = 0$ and $\lambda_2^{(\mathrm{i})} = 0$,
with the additional constraint $\cos\alpha > 0$.
The former equation is equivalent to
\begin{equation}
1 + \sqrt{1 - \chi^2} + \chi \tan\alpha = 0.
\label{Eq:lambda:1}
\end{equation}
Rearranging the terms and squaring both sides of the new equation, we obtain
$$
1 - \chi^2 = ( 1 + \chi \tan\alpha )^2,
$$
or
$$
\frac{\chi ( \chi + \sin 2\alpha )}{\cos^2\alpha} = 0.
$$
It is easy to verify that $\chi = 0$
is not a solution of Eq.~(\ref{Eq:lambda:1})
and $\chi = - \sin 2\alpha$ satisfies Eq.~(\ref{Eq:lambda:1}) but only for $\cos 2\alpha < 0$.
Similarly, we can analyze the equation $\lambda_2^{(\mathrm{i})} = 0$ and obtain the second
stability boundary $\chi = \sin 2\alpha$.
Together with formula~(\ref{Def:chi}),
this gives the following stability boundaries
of synchronous in-phase cluster states
$$
\delta_\mathrm{cr}^{(\mathrm{i})} = \pm \e \cos\alpha \sin 2\alpha
\quad\mbox{for}\quad\pi/4 < \alpha < \pi/2.
$$

\subsection{Asynchronous state}

The integrals in Eq.~\eqref{lambda_running} can be found analytically. For this, we rewrite
formulas~\eqref{lambda_running} as
\begin{eqnarray*}
\lambda_1 &=& \left( - \frac{\e}{2 C} \right) \int_0^{2\pi} \frac{\cos\alpha + \cos(\psi-\alpha)}{| \chi - \sin \psi |} \dd\psi
= - \frac{\e}{2} \left( \cos\alpha + \frac{\sin\alpha}{C} \int_0^{2\pi} \frac{\sin\psi\: \dd\psi}{| \chi - \sin \psi |} \right),
\\[2mm]
\lambda_2 &=& \left( - \frac{\e}{2 C} \right) \int_0^{2\pi} \frac{\cos\alpha + \cos(\psi+\alpha)}{| \chi - \sin \psi |} \dd\psi
= - \frac{\e}{2} \left( \cos\alpha - \frac{\sin\alpha}{C} \int_0^{2\pi} \frac{\sin\psi\: \dd\psi}{| \chi - \sin \psi |} \right).
\end{eqnarray*}
where
$$
C = \int_0^{2\pi} \frac{\dd\psi}{| \chi - \sin \psi |} = \frac{2\pi}{\sqrt{\chi^2 - 1}}.
$$
Then, it is easy to verify that
$$
\int_0^{2\pi} \frac{\sin\psi\: \dd\psi}{| \chi - \sin \psi |} = - 2 \pi\: \mathrm{sign}\:\chi + \chi C.
$$
Therefore,
\begin{eqnarray*}
\lambda_1 &=& - \frac{\e}{2} \left( \cos\alpha + \left( \chi - \sqrt{\chi^2 - 1}\: \mathrm{sign}\:\chi \right) \sin\alpha \right),
\\[2mm]
\lambda_2 &=& - \frac{\e}{2} \left( \cos\alpha - \left( \chi - \sqrt{\chi^2 - 1}\: \mathrm{sign}\:\chi \right) \sin\alpha \right).
\end{eqnarray*}

In this case, equations $\lambda_1 = 0$ and $\lambda_2 = 0$ can also be solved analytically.
In particular, the former equation has a solution
$$
\chi = - \frac{1}{\sin 2\alpha}
\quad\mbox{for}\quad
\pi/4 < \alpha < \pi/2,
$$
while the latter equation has a solution with the opposite sign.
Using the definition of $\chi$,
we conclude that an asynchronous cluster state
loses its transversal stability
for the critical frequency detuning
$$
\delta_\mathrm{cr} = \pm \frac{\e}{2 \sin\alpha}.
$$

\section{Performing the variable transformation in the three-cluster state}
\label{App2}

For the $(2+1)$ state, let us rewrite the system~(\ref{eqn:2+1_a})-(\ref{eqn:2+1_c}),
using a variable transformation
$(\Theta_1,\Theta_2,\Phi_1)\rightarrow (\Xi,\Psi,\Phi)$, where
$$
\Xi = \frac{\Theta_1+\Theta_2}{2},\quad \Psi = \frac{\Theta_1-\Theta_2}{2}, \quad \Phi=\Theta_1-\Phi_1,
$$
or inversely
$$
\Theta_1=\Xi+\Psi,\quad \Theta_2=\Xi-\Psi,\quad \Phi_1=\Xi+\Psi-\Phi.
$$
By adding or subtracting two equations
from the system~(\ref{eqn:2+1_a})-(\ref{eqn:2+1_c}), we 
obtain:
\begin{eqnarray}   
    \dot{\Xi} &=& \Omega_1 + \mathrm{Im}\left[f_1\left(Z^{(1,2)},Z_2^{(1,2)}; \e, \alpha, \kappa\right)
    \cos{\Psi}\cdot\ee^{-\ii\Xi} + f_2\left(Z^{(1,2)},Z_2^{(1,2)}; \e, \alpha, \kappa\right)
    \cos(2\Psi)\cdot\ee^{-2\ii\Xi} \right]\,,\label{Eq:Xi:App}\\
    \dot{\Psi} &=& 
    -\mathrm{Re}\left[f_1\left(Z^{(1,2)},Z_2^{(1,2)}; \e, \alpha, \kappa\right)\sin\Psi\cdot\ee^{-\ii\Xi} 
    + f_2\left(Z^{(1,2)},Z_2^{(1,2)}; \e, \alpha, \kappa\right)\sin(2\Psi)\cdot\ee^{-2\ii\Xi} \right]\,,\label{Eq:Psi:App}\\
    \dot{\Phi} &=& \omega_1-\omega_2 + \mathrm{Im}\left[f_1\left(Z^{(1,2)},Z_2^{(1,2)}; \e, \alpha, \kappa\right)
    (1-\ee^{\ii\Phi}) \ee^{-\ii\Psi}\cdot \ee^{-\ii\Xi}\right. \nonumber\\
    &+& \left.f_2\left(Z^{(1,2)},Z_2^{(1,2)}; \e, \alpha, \kappa\right)
    (1-\ee^{2\ii\Phi})\ee^{-2\ii\Psi}\cdot\ee^{-2\ii\Xi} \right]\,.\label{Eq:Phi:App}
\end{eqnarray}
The explicit form of the functions $f_1$ and $f_2$
is given by the formulas~\eqref{Def:f_1}--\eqref{Def:xi_1:xi_2}, where the variables $\xi_1$ and $\xi_2$ are polynomial functions
of the mean fields $Z^{(1,2)}$ and $Z_2^{(1,2)}$.
Using Eq.~\eqref{Def:xi_1:xi_2}, we obtain:
\begin{eqnarray*}
    \xi_1 &=& Z^{(1)}+Z^{(2)}=\left(\cos\Psi+\ee^{\ii(\Psi-\Phi)}\right)\cdot\ee^{\ii\Xi}\,,\\
    \xi_2 &=& \left(\overline{Z}^{(1)}+\overline{Z}^{(2)}\right)\left(Z_2^{(1)}+Z_2^{(2)}\right)=
    \left(\cos\Psi+\ee^{-\ii(\Psi-\Phi)}\right)
    \left(\cos2\Psi+\ee^{2\ii(\Psi-\Phi)}\right)\cdot\ee^{\ii\Xi}\,,\\
    \xi_1^2 &=& \left(\cos\Psi+\ee^{\ii(\Psi-\Phi)}\right)^2\cdot\ee^{2\ii\Xi}\,.
\end{eqnarray*}
This allows us to rewrite the functions $f_1$ and $f_2$ as
$$
    f_1\left(Z^{(1,2)},Z_2^{(1,2)}\right) = g_1(\Psi,\Phi;\e,\alpha,\kappa)\cdot\ee^{\ii\Xi}\,,\quad
    f_2\left(Z^{(1,2)},Z_2^{(1,2)}\right) = g_2(\Psi,\Phi;\e,\alpha,\kappa)\cdot\ee^{2\ii\Xi}\,,
$$
where
\begin{eqnarray}
    g_1(\Psi,\Phi;\e,\alpha,\kappa) &=& \frac{\e}{2}\ee^{\ii\alpha}\left(1-\frac{\e}
    {2\kappa}\ee^{\ii\alpha}\right)\left(\cos\Psi+\ee^{\ii(\Psi-\Phi)}\right)
    - \frac{\e^2}{8\kappa}\left(\cos\Psi+\ee^{-\ii(\Psi-\Phi)}\right)
    \left(\cos2\Psi+\ee^{2\ii(\Psi-\Phi)}\right)\,, \label{Def:g_1}\\
    g_2(\Psi,\Phi;\e,\alpha,\kappa) &=& \frac{\e^2}{8\kappa} \ee^{2\ii\alpha}
    \left(\cos\Psi+\ee^{\ii(\Psi-\Phi)}\right)^2\,. \label{Def:g_2}
\end{eqnarray}
Inserting these expressions into Eqs.~\eqref{Eq:Xi:App}--\eqref{Eq:Phi:App}, we find that all the terms with $\Xi$ cancel out.
Thus, we obtain an equivalent system
\begin{eqnarray}
    \dot{\Xi} &=& \Omega_1 + \mathrm{Im}\left[g_1(\Psi,\Phi;\e,\alpha,\kappa)\cos{\Psi} 
    + g_2(\Psi,\Phi;\e,\alpha,\kappa)\cos(2\Psi) \right]\,,\label{eqn:2+1_alt_a}\\
    \dot{\Psi} &=& 
    -\mathrm{Re}\left[g_1(\Psi,\Phi;\e,\alpha,\kappa)\sin\Psi
    + g_2(\Psi,\Phi;\e,\alpha,\kappa)\sin(2\Psi) \right]\,,\label{eqn:2+1_alt_b}\\
    \dot{\Phi} &=& \omega_1-\omega_2 + \mathrm{Im}\left[g_1(\Psi,\Phi;\e,\alpha,\kappa)
    (1-\ee^{\ii\Phi}) \ee^{-\ii\Psi} 
    + g_2(\Psi,\Phi;\e,\alpha,\kappa)\cdot(1-\ee^{2\ii\Phi})\ee^{-2\ii\Psi} \right]\,.\label{eqn:2+1_alt_c}
\end{eqnarray}
Note that the r.h.s. of all three equations depend only on two variables $\Psi$ and $\Phi$,
which means that the dynamics of the three-cluster state is essentially two-dimensional.
More precisely, it is dynamics on a two-dimensional torus
parameterized by variables $\Psi$ and $\Phi$.
The third variable $\Xi$ is simply a slave variable,
since the r.h.s. of Eq.~\eqref{eqn:2+1_alt_a} does not depend on $\Xi$.

%

\end{document}